%% file: superb-review.tex
\documentclass[12pt]{iopart}

\usepackage{xspace}  
\usepackage{graphicx}

\input{superbsym.tex}

\begin{document}

\topical[\boldmath The Physics of Heavy Flavours at \superb]
{The Physics of Heavy Flavours at \superb}

\author{Dr Adrian J. Bevan}

\address{Department of Physics,
Queen Mary, University of London,
Mile End Road,
London, E1 4NS, UK}
\ead{a.j.bevan@qmul.ac.uk}
\begin{abstract}
This is a review of the \superb project, covering the accelerator, detector, and 
highlights of the broad physics programme.
\superb is a flavour factory capable of performing precision measurements
and searches for rare and forbidden decays of $B_{u,d,s}$, $D$, $\tau$ and $\Upsilon({\mathrm{nS}})$
particles.  These results can be used to test fundamental symmetries 
and expectations of the Standard Model, and to constrain many different
hypothesised types of new physics.  In some cases these measurements can be
used to place constraints on the existence of light dark matter and 
light Higgs particles with masses below 10\gevcc.
The potential impact of the measurements that will be made by \superb on the field
of high energy physics is also discussed in the context of data taken
at both high energy in the region around the \FourS, and near charm threshold.
\end{abstract}

\maketitle

\input{intro}

\input{facility}
\input{tau}

\input{bphysics}
\input{dphysics}

\input{precisionew}

\input{directsearches}

\section{Other measurements}
\label{other}

There is a vast potential to perform other measurements that are not classified as either 
a {\em golden mode}, precision CKM, or SM measurement.  These other measurements include hundreds of 
possible decays of $B$, $D$, $\tau$, $\Upsilon$, and $\psi(3770)$ as well as 
studies of initial state radiation processes, and both conventional and exotic spectroscopy not discussed here.
It should be noted that the physics programme at other resonances above the $\psi(3770)$ is
under study, and has not been discussed here.
A partial description of many of these possibilities can be found in Refs.~\cite{physicswp,Bona:2007qt,Hitlin:2008gf},
and a more comprehensive summary will be discussed in the context of results of existing experiments 
in the forthcoming Physics of the B Factories book currently in preparation~\cite{PBF}.  \superb will 
integrate 150 (75) times the data of \babar and \belle enabling a significant improvement in precision of 
these other measurements, and it is expected that during the lifetime of this experiment a number of 
new areas will be developed.

\input{interplay}

\input{summary}

\ack
\input{ack}

\section*{References}

\bibliographystyle{unsrt}
\bibliography{superb-review}

\end{document}

%% file: intro.tex
%
%
\section{Introduction}

This topical review discusses the potential impact of 
high luminosity \epem collider experiments, the so-called Super 
Flavour Factories, on our understanding of high energy physics.  
In particular this review will focus on the potential of one of 
these facilities; the \superb project. This
experiment will record billions 
of $B$, $D$, and $\tau$ decays at various center of mass energies ranging 
between the $\psi(3770)$ and $\Upsilon(6S)$ in order to search for signs of 
physics beyond the Standard Model of Particle Physics (SM) and to perform 
precision measurements of the SM.
There are two Super Flavour Factories, one called \belletwo which is being 
constructed at KEK in Japan, and the other called \superb being built in Italy.
The aim of \belletwo is to integrate 50\invab of data at a center of mass energy
corresponding to the \FourS resonance, while
\superb aims to integrate 75\invab of data at that energy.
There are several important differences between these two experiments (i)
the electron beam at \superb will be polarised, enabling superior performance in
the study of $\tau$ leptons and other important precision tests of the SM such as
the measurement of the weak mixing angle via
$\sin^2\theta_W$ and (ii) \superb will have a dedicated run at 
the $\psi(3770)$ which corresponds to the charm production threshold. Before 
discussing the implications of the many measurements that will be
possible at \superb, and thus the benefits of the additional features of \superb over
\belletwo, it is prudent to take a brief look at history (Section~\ref{intro:history}), and our 
understanding of current popular expectations of physics beyond the SM, 
which is often referred to a `{\em new physics}' (NP) in the literature (Section~\ref{intro:beyondthesm}).
Section~\ref{intro:outline} provides an outline of the rest of the review.
Detailed reviews of the physics programmes of \superb and \belletwo can be found in
Refs.~\cite{physicswp} and~\cite{belleii}.

There are two types of measurement that provide the motivation for \superb.
The first type consists of theoretically clean observables that can be measured
with high precision.  Such observables for a rare or suppressed decay can 
be sensitive probes of NP.  Decay channels related to this type of 
measurement are often referred to as {\em golden modes} in the context of NP searches.  
The second type of measurement motivating the \superb experiment are precision CKM or 
SM measurements, for example the precision measurement of $\sin^2\theta_W$.    These
measurements have a dual purpose, to provide a precision determination of SM
parameters, and in turn to constrain possible NP scenarios.

\subsection{Historical look at flavour}
\label{intro:history}

The SM provides a mathematical description of all known physical phenomena relating to 
the interactions between particles and anti-particles.  Where the particles are divided into
quarks (up-type quarks are $u$, $c$, and $t$, and the down-type quarks are $d$, $s$, and $b$), 
leptons ($e$, $\mu$, $\tau$ and their respective 
neutrinos) and gauge bosons ($\gamma$, $g$, $Z^0$, $W^\pm$).  The sub-set of phenomena relating
to the change of one type of quark or lepton into another type of quark or lepton is referred
to as {\em flavour} physics.  Phenomena pertaining to flavour interactions in quarks are described
by the $3 \times 3$ Cabibbo-Kobayashi-Maskawa (CKM) quark mixing matrix, which encompasses 
the quark-mixing mechanism postulated by Cabibbo in 1963~\cite{Cabibbo:1963yz} with the description of \CP violation 
introduced by Kobayashi and Maskawa in 1973~\cite{Kobayashi:1973fv}.  In the SM Lagrangian transitions of up-type ($q_u$) 
and down-type ($q_d$) quarks are mediated by the exchange of a $W$ boson via
$q_u V_{ij} q_d$,
where $V_{ij}$ represents the CKM matrix with the indices $i$ and $j$ corresponding 
to the quarks.  The CKM matrix is
\begin{eqnarray}
\vckm & =& \theckmmatrix,\\
      &\simeq &\theckmmatrixwolf. \label{bevan:eq:ckm}
\end{eqnarray}
The expansion shown above is in terms of $\lambda$, the sine of the Cabibbo angle, and 
three other parameters $A$, $\rho$ and $\eta$.  This description of the CKM matrix
is the convention of Wolfenstein~\cite{Wolfenstein:1983yz}.  When working with the large
data samples expected at the next generation of experiments, the above matrix
is not expanded to sufficient orders in $\lambda$, and one obtains a convention dependent
solution.  The Buras parameterisation of the CKM matrix provides a convenient framework
to use at higher orders~\cite{Buras:1994ec,Bevan:2011up}.
\superb is able to probe in detail many aspects of the CKM matrix by making a 
number of redundant measurements.

There is an equivalent formalism to describe
neutrino mixing that is currently being explored by a number of experiments
including Daya Bay, Super Kamiokande, and T2K.  These measurements are related to elements in a 
$3\times 3$ mixing matrix describing neutrino mixing (for example see the review by B.~Kayser 
in Ref~\cite{Nakamura:2010zzi}).
To date there is no evidence for charged lepton flavour violation.
\superb will be able to make significant advances in the search for charged lepton 
flavour violation in $\tau$ decays, so transitions from the third, to the second 
or first generations of charged leptons (see Section~\ref{tau}).  

Historically {\em quark mixing} was postulated as a way to understand the behavior of hadronic 
currents in Hyperon decays.  This work was completed in an era before the concept of 
quarks was accepted as a given fact, and the concept of mixing was expressed in terms of currents,
that would have corresponded to interactions between the $u$, $d$, and $s$ quarks in today's 
terminology. Shortly after this significant step forward, it was realised that attempts to 
reconcile theory and measurement for the branching fraction of $K_L^0 \to \mu^+\mu^-$ decays 
required the introduction of a fourth quark via the GIM mechanism~\cite{Glashow:1970gm}.  
The discovery
of the $J/\psi$ particle was the confirmation that this fourth quark existed, and 
we now refer to this as the charm quark.  
A repeat of this problem was encountered in the study of $\Bz-\Bzb$ mixing by
the ARGUS experiment.  The amplitude for $\Bz-\Bzb$ mixing is dominated
by transitions involving the top quark, thus theorists were able to make 
predictions of the top quark mass based on the experimental knowledge of this
mixing observable.  It is interesting to note that in both cases (i) the study of 
a rare kaon decay, leading to the discovery of the charm quark, and (ii)
the use of experimental constraints on $\Bz-\Bzb$ mixing to discover the 
top quark, one is using a low energy flavour changing process to place stringent
constraints on a much higher energy phenomenon.  
Measurements of flavour changing processes
such as the GIM mechanism and mixing in $B^0\overline{B}^0$ decays have shaped
our understanding of the SM.  

Precision measurements in the flavour sector will continue to provide a detailed
set of reference points to test models of NP against.  
This aspect underpins the importance of many of the measurements that will be 
made at \superb.
In addition to these
particle physics constraints, there are also ramifications for other fields of 
research such as astrophysics, in terms of searches for Dark Matter candidates,
and ultimately cosmology in terms of understanding the evolution of matter and
anti-matter in the early universe.

\subsection{Expectations for physics beyond the Standard Model}
\label{intro:beyondthesm}

The experimental community has been focusing on the search for evidence of 
the Higgs particle, which would be added to the Standard Model (SM) in order
to make this more self-consistent.  Having introduced the SM Higgs to the model,
further corrections are required in order to cope with Higgs self coupling interactions.
This motivates the search for a richer texture of NP beyond just identifying a 
SM Higgs candidate.  If it turned out that the Higgs did not exist, then something 
else would have to be introduced into the model in order to address the issues
that the Higgs particle was originally postulated for. There is a wide range
of scenarios of physics beyond the SM that have been postulated.  Many of these
scenarios are derived from some higher theory such as M-Theory or sub-sets such as
SUSY, others introduce a variety of different concepts, for example extra spatial
dimensions, additional generations of fermions, and additional Higgs particles.  
These models of new physics
are obtained by adding new terms to the SM Lagrangian, and then using existing
constraints from experiments to evaluate if such an addition is consistent with 
nature or not.  Some of the most stringent constraints that 
have guided theorists in the construction of the SM are so-called flavour changing
neutral currents (FCNC), and in many cases such constraints are being used to guide the
development of theories beyond the SM. 

The criteria required to
probe the high energy regime are to identify suppressed processes within the
SM that are theoretically clean that may have contributions from new heavy particles,
and then to perform precision measurements of those processes.  
Interpretation of the results, in comparison with both SM expectations,
and those of the NP scenario can be used to constrain the parameter space
of the NP model.  One of the parameters that enters into this process is the 
energy scale for the new physics $\Lambda_{NP}$ (e.g. the particle mass in
the case of the charm and top quarks discussed previously).
Thus just as flavour changing processes have provided stringent constraints for theorists 
in understanding and constructing the SM, any theory of physics that goes
beyond the scope of the SM will also be strongly constrained by measurements of flavour 
changing transitions.  
Hence model builders will be able to partially reconstruct the new physics Lagrangian 
using the results of \superb and other flavour experiments.  In particular if there
are mixing matrices between sets of new particles introduced into the theory,
then in general the off-diagonal complex elements may be constrained by 
rare decays probed in flavour physics experiments.

\subsection{The outline of this review}
\label{intro:outline}

The remainder of this review paper provides a description of the 
experimental facility (Section~\ref{sec:facility}), followed by a 
pedagogical overview of the physics potential of \superb.  
The following sections discuss the roles of 
$\tau$ physics in terms of searches for forbidden
processes that violate well known symmetries of nature, and tests of the SM (Section~\ref{tau}),
the decays of $B_{u,d,s}$ (Section~\ref{bphysics}),
$D$ mesons (Section~\ref{dphysics}), and precision 
tests of electroweak physics which are discussed in Section~\ref{precisionew}.  
Spectroscopy measurements in terms of direct searches of unknown particles related to new physics 
(Section~\ref{directsearches}) is also reviewed.
Section~\ref{other} briefly mentions some of the other measurements that can be made at \sffs.
Estimated improvements in the field of Lattice QCD and subsequent impact on the \superb
physics programme are discussed in Ref.~\cite{physicswp}.  
Thus far the \lhc experiments have not found evidence for the SM Higgs or any physics 
beyond the SM.  The exclusions obtained using the first few years of data taking
suggest that flavour observables can, and will, play an important role in elucidating
nature in the coming years.  It is important to globally combine information from
all possible measurements together in order to optimally decode the signatures of physics
beyond the SM.  Section~\ref{interplay} reviews the measurements to be made at \superb
in such a global context, highlighting inter-relations between different sets of measurements
as a tool to elucidate generic behavior of new physics and of the SM.

%% file: facility.tex
\section{The \superb experimental facility}
\label{sec:facility}

The most up to date detailed review of the \superb experimental programme is available in the form
of a set of {\em progress reports} discussing the accelerator~\cite{acceleratorwp}, detector~\cite{detectorwp}, 
and physics~\cite{physicswp,Meadows:2011bk}.  
Older descriptions of the project can be found in Refs.~\cite{Bona:2007qt,Hitlin:2008gf}.
The \superb collaboration is also in the process of preparing a set of Technical Design 
Reports that will supersede these reports and serve as blue prints for the construction of the 
experiment.
This section provides a brief summary of the aspects of \superb accelerator and detector
as detailed in the aforementioned reports.  
\superb will be constructed an the Cabibbo Laboratory, Tor Vergata
University near Rome, Italy. First collisions could be as early as 2016, 
with the first year of nominal data taking starting the following year,
After five years of nominal data taking this experiment
should have integrated 75\invab of data at the \FourS, which is 150 times that of \babar,
and 75 times that of \belle.  On a similar time-scale (early next decade) 
the competing experiment \belletwo
will have accumulated 50\invab of data.

By the time \superb starts taking data, the \lhcb experiment will have finished much of its 
physics programme. Any discoveries of new physics from \lhcb, or indeed possible inconsistencies
of measurement and SM expectations will be of direct interest to the \superb physics programme.
This is the case as, aside from FCNC and annihilation topologies, the difference between $B_{u,d}$ 
decays and $B_{s}$ decays is the choice of the light spectator quark, which does not drive 
the physics content of a particular decay. Any FCNC or annihilation topologies that do manifest 
signs of a deviation from the SM in $B_s$ decays also have parallels for $B_{u,d}$ decays, although
the relative new physics couplings may in general be different from the $B_s$ case.
While hadron collider experiments have the advantage of vast statistics
over experiments at $\epem$ machines, there is a price paid in terms of triggering systems, 
backgrounds, and poor neutral reconstruction.
Typically rare charged hadronic final states will be measured well in hadron machines, and hence 
good theoretical control of hadronic uncertainties may be required to interpret such results.
Whereas experiments at $\epem$ will excel in final states containing neutrals, and in particular $\nu's$ 
that would otherwise be challenging or impossible to study in a hadronic environment.
Final states of this type are generally theoretically much cleaner that the hadronic ones best accessed
in a hadron machine.  Many of 
these latter decays are vital ingredients for constraining possible sources of NP.
A proposal for a potential upgrade of \lhcb is in preparation, and such an 
experiment could finish taking data as early as 2030~\cite{lhcbupgradeloi}, a number of 
years after \superb and \belletwo are expected to have finished accumulating their nominal 
data sets. 

\subsection{The accelerator}
\label{sec:facility:accelerator}

The \superb accelerator is designed to collide bunches of electrons and positrons 
at center of mass energies between 3.37\gev and 11\gev, such that the center of mass system
is boosted in the reference frame of the laboratory.  The reason for having a boosted center
of mass frame is in order to facilitate the study of time-dependent \CP violation in $B$ and $D$ decays, 
and this naturally results in a forward-backward asymmetry in the design of the 
detector.  This requirement has a consequence that asymmetric beam energies are needed,
with a boost factor of the centre of mass relative to the lab $\beta\gamma = 0.23$ at the \FourS. 
While the baseline design for the machine also has a low boost factor for operation at
the $\psi(3770)$, it may be possible to take data at this energy with a with a boost factor $\beta\gamma$
as large as 0.91.  If that were realisable, then this would open up the possibility of performing
a number of quantum correlated time-dependent studies with charm decays as discussed in 
Section~\ref{dphysics}.

The instantaneous luminosity of the accelerator at the \FourS [$\psi(3770)$] will be 
$10^{36} [10^{35}] cm^{-2}s^{-1}$, with 
bunch currents of a few amps in both the low and high energy rings (LER and HER).  This is an 
increase of two orders of magnitude in luminosity compared to the operating
conditions at PEP II, with similar beam currents in both rings.
The luminosity in a circular collider is given by
\begin{eqnarray}
{\cal L} =\frac{N^+N^- f_c}{4\pi \sigma_y \sqrt{[\sigma_z \tan(\theta/2)]^2+\sigma_x^2}}
\end{eqnarray}
where $N^{\pm}$ are the number of electrons ($-$) and positrons ($+$) in a bunch, $\sigma_{i}$ is
the beam size in dimension $i=x, y, z$ where $\sigma_{x, y} = \sqrt{\beta^*_{x, y}\epsilon_{x,y}}$.  Here 
$\beta^*_i$ is the beta function and $\epsilon_i$ is the emittance at the interaction point (IP).
The angle $\theta$ is the crossing angle of the two beams, so the $\sigma_z$ contribution 
vanishes for beams colliding head on, and the parameter $f_c$ is the frequency of collision of each bunch.
There are two potential routes to increasing the luminosity of a circular collider (i) increase
the number of electrons (which is related to the power required to run the machine), or
(ii) decrease the transverse size of the bunches in the beam, i.e. decrease the emittance.
The main driving force to increasing the luminosity of the \superb accelerator
relative to PEP II is to significantly reduce the emittance of the beam, and as a result to
make the bunch sizes smaller than the previous generation of \epem colliders.  There is 
a second important improvement in the accelerator design related to the interaction region:
In order to bring bunches of electrons and positrons into collision, one has to either
have a complicated array of magnets to align an incoming $e^-$ bunch so that it collides
head on with a $e^+$ bunch, or to have a small but finite crossing angle between the 
two beams.  The former approach was adopted by the SLAC \BF, and has the limitation that
there would be a significant level of luminosity related beam backgrounds recorded 
in the \superb detector that could obscure some of the rare signals under study if a similar 
approach was adopted for future machines.  The
traditional problem with the latter approach results from the fact that bunches
of electrons and positrons are ellipsoidal in shape, and by bringing two 
bunches into collision at a finite angle, and the effective cross sectional area of 
the collision at the IP is reduced.  It was realised by Pantaleo Raimondi 
that sextupole magnets positioned before and after the interaction region can be 
used to skew the transverse waists of incoming bunches with a finite crossing angle
in such a way that they are pinched optimally at the point of collision.  This 
collision scheme has been termed the `crabbed waist' scheme, and it was successfully
tested at LNF Frascati in 2009 (See ~\cite{acceleratorwp} and references therein).

A unique feature of the machine is that the bunches of electrons will be polarised,
which translates into significant benefits for the \superb physics programme.
The polarisation is designed to be $\sim 80\%$.  Two particular benefits of this feature are
(i) this provides an additional kinematic variable in studying 
rare $\tau$ decays and is useful for both $\tau$ EDM and $g-2$ measurements, 
as one is able to reconstruct the polarisation of the $\tau$ in the final state and 
use this as a background suppression tool, and (ii) one can perform precision electroweak tests 
of the SM, such as measuring $\sin^2\theta_W$ using left-right asymmetries in 
$e^+e^-\to f\overline{f}$ transitions, where $f$ is a fermion, in addition to 
being able to measure forward-backward asymmetries that would be accessible without
a polarised beam.  The asymmetry for the 
$b\overline{b}$ final state can be measured as precisely as the SLC/LEP measurements 
at the $Z$ pole, but at an energy that is free from hadronisation uncertainties. Details of the intended
scheme to be used in order to obtain a polarised electron beam can
be found in Ref.~\cite{acceleratorwp}.  In order to use the
polarisation information in precision measurements one needs to have a sub $1\%$ measurement of
the value of the polarisation, which is achievable using a Compton polarimeter.

A number of parameter sets have been developed for use at \superb for both operation
at a center of mass energy corresponding to the \FourS, and to operate at the 
$\psi(3770)$.  In addition to being able to operate the machine at these two resonances,
it will be possible to scan the machine from the \OneS to the \SixS resonance.
The physics programme for prolonged running at one of the other $\Upsilon$ resonances
rests in direct searches for light dark matter and Higgs particles, tests
of lepton flavour universality, and the study of $B_s$ mesons in the case of the \FiveS.
Table~\ref{tbl:machineparameters} shows some of the parameters for the different configurations of the 
accelerator.  All of the three \FourS configurations are able to reach the
desired luminosity of $10^{36} cm^{-2}s^{-1}$, and while the collection of 
machine parameters may look difficult to achieve, each of these has been 
demonstrated at an operating machine somewhere in the world.  The challenge
on the accelerator side is to construct a machine that can simultaneously 
achieve all of these. The $\psi(3770)$ configuration should reach an instantaneous
luminosity of $10^{35} cm^{-2}s^{-1}$.  All other resonances of interest near the
\FourS and  $\psi(3770)$ energies can be reached by tuning the machine lattice from
one of the two optimal working points. A more detailed description of the \superb accelerator 
can be found in Ref.~\cite{acceleratorwp}.

\begin{table}[!h]
\caption{Machine parameters for different configurations of the accelerator.
There are three configurations for operating at the \FourS, and one for the $\psi(3770)$.}
\label{tbl:machineparameters}
\begin{center}
\begin{tabular}{l|cccc}\hline
Parameter                  & Nominal & Low emittance&High Current    & $\psi(3770)$ \\ \hline
$e^-$ / LER energy (\gev)          & 4.18    & 4.18         &    4.18        & 1.61 \\
$e^+$ / HER energy (\gev)          & 6.7     & 6.7          &    6.7         & 2.58\\
$\epsilon_y$ (HER) (pm)    & 5.0     & 2.5          &    10          & 13\\
$\epsilon_x$ (HER) (nm)    & 2.0     & 1.0          & 2.0            & 5.2\\
$\epsilon_y$ (LER) (pm)    & 6.15    & 3.08         & 12.3           & 16\\
$\epsilon_x$ (LER) (nm)    & 2.46    & 1.23         & 2.46           & 6.4\\
$\sigma_y$ (HER) ($\mu m$) & 0.036   & 0.021        & 0.054          & 0.092\\
$\sigma_x$ effective (HER) ($\mu m$) & 165.22  & 165.22       & 145.60         & 166.67\\
$\sigma_y$ (LER) ($\mu m$) & 0.036   & 0.021        & 0.0254         & 0.092\\
$\sigma_x$ effective (LER) ($\mu m$) & 165.30  & 165.30       & 145.78         & 166.67\\
Total Power (MW)           & 16.38   & 12.37        & 28.83          & 2.81 \\ 
$e^-$ Polarisation (\%)    & 80.0    & 80.0         & 80.0 & $-$ \\ \hline
${\cal L}$ ($cm^{-2}s^{-1}$) & $10^{36}$ & $10^{36}$ & $10^{36}$ & $10^{35}$ \\
 \hline
\end{tabular}
\end{center}
\end{table}

While the primary goal of \superb is the pursuit of knowledge that will hopefully elucidate our
understanding of physics beyond the SM, it has also been realised that the small emittance of the 
\superb machine means that this facility will be an extremely bright synchrotron light source.
In fact this machine will be thirty times brighter than the Diamond facility at RAL, UK and the ESRF facility
at Grenoble, France.  While this is an interesting subject in itself, it is not the focus
of this review, and will not be discussed further here.

\subsection{The detector}

\begin{figure}[!ht]
\begin{center}
\resizebox{14.cm}{!}{
\includegraphics{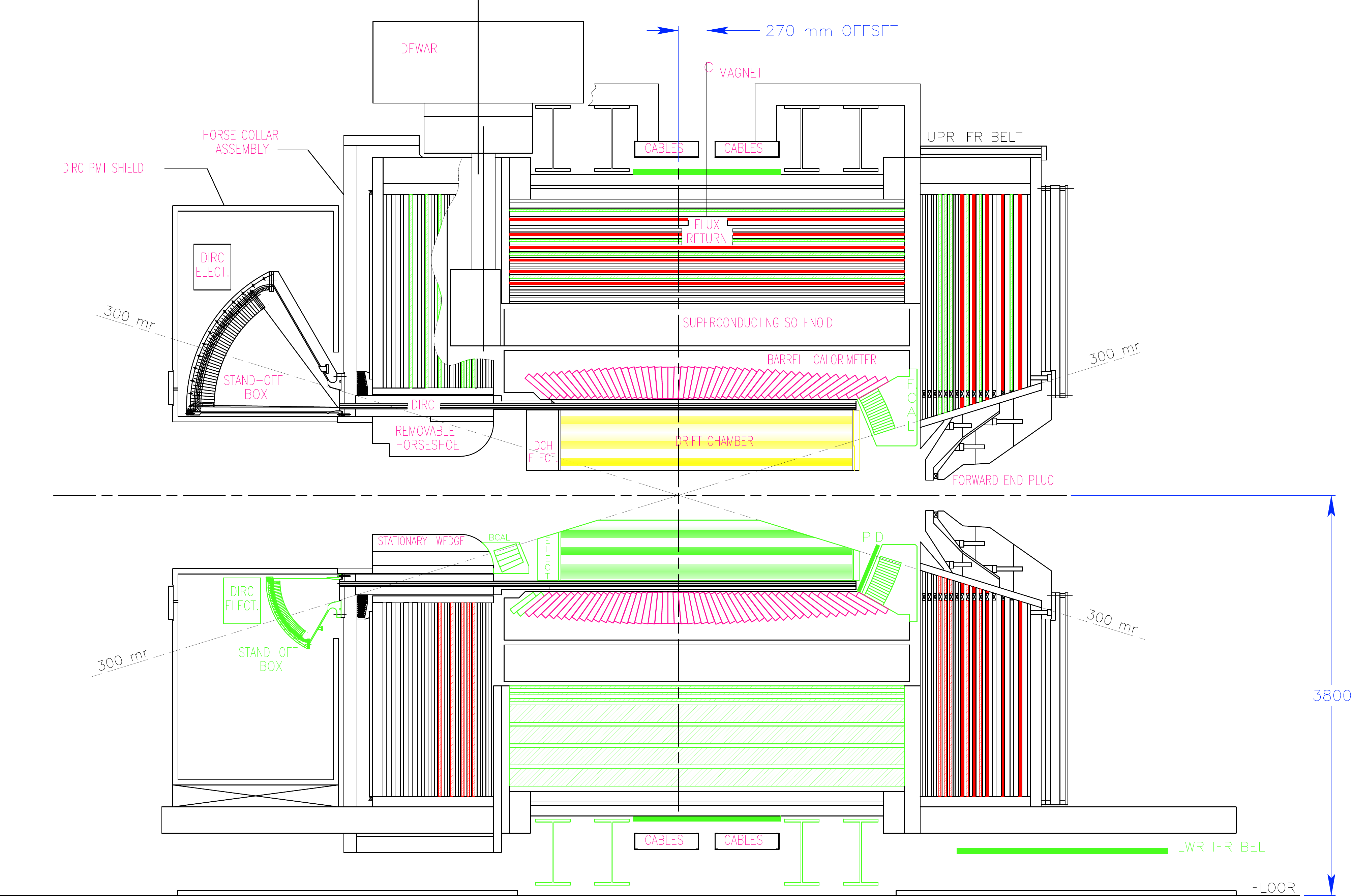}
}
\caption{A schematic of the \superb detector concept.}\label{fig:detector}
\end{center}
\end{figure}

Working from the inner to outermost components, the \superb detector consists of a Silicon Vertex 
Tracker (SVT) surrounded by a Drift Chamber (DCH), both of which are used to detect the passage of 
charged particles through the detector.  The Particle Identification (PID)
system is comprised of 
a next generation Detector of Internally Reflected Cherenkov radiation  (FDIRC) 
which surrounds the DCH.  There is also a forward PID system under investigation
to provide particle identification over a larger solid angle than that covered by the FDIRC.
Surrounding the PID system is an Electromagnetic Calorimeter (EMC) which is used primarily to
provide measurements of photon and electron energies. All of the aforementioned
components will be situated in a super-conducting solenoid magnet capable of producing a solenoidal 
magnetic field of 1.5T.  This is the same solenoid magnet that was used for the \babar experiment.
The field strength for charm threshold running may be lower than 1.5T, and studies are
ongoing in order to determine the strength required for signal reconstruction versus background
suppression.
The outermost part of the detector is the so-called Instrumented Flux Return (IFR) which is used to  
identify muons and $\KL$ mesons.  Figure~\ref{fig:detector}, taken from Ref.~\cite{detectorwp},
 shows a schematic of the \superb detector.  The top half of the Figure illustrates the baseline
detector design, while the bottom half shows various options such as forward PID system and 
backward calorimeter. The main components of the detector are described in 
more detail in the following (A more detailed discussion can be found in Ref.~\cite{detectorwp}).

\begin{description}
  \item{\bf SVT}:
The \superb SVT consists of two types of detector.  The first is a highly segmented device
close to the beam pipe (referred to as Layer 0), and the second is a \babar-like multi-layer 
double sided silicon strip detector.  There are several choices of technology for Layer 0, 
and the baseline choice for low luminosity operation of the machine (data taking
during the first few years) is a double-sided silicon strip based detector with short 
strips at $45^\circ$ to the direction of the beam, with a stereo angle of
$90^\circ$ between strips on both sides of the sensor.
This iteration of Layer 0 will be at an average radius of 1.6cm from the IP.  There are
several Monolithic Active Pixel Sensor and and Hybrid Pixel Sensor technology options
under investigation, all of which will be suitable for use in the high luminosity running 
that corresponds to nominal data taking.   The amount of material in Layer 0 depends on the 
technology chosen, but is expected to be between 0.4 and 1.0\% of a radiation length.
The outer part of the SVT will consist of 
several layers of double sided strip sensors arranged in a configuration similar to the \babar 
SVT~\cite{Aubert:2001tu}.  The use of the SVT is not anticipated in the trigger, and triggers
formed using information from the DCH will be required in order to read out the SVT.

  \item{\bf DCH}:
The \superb drift chamber will be the primary sub-system for providing measurement information
on tracks with momenta larger than $\sim 100\mevc$.  The DCH design for \superb is similar to the \babar one, with 
40 layers of cells, the cross-section of each cell being $1cm^2$.  Studies are underway to 
determine the optimal gas mixture to use for the DCH, and the overall layout, with either spherical
or stepped end plates.  The choice of gas mixture will be driven by the desire to minimize the 
overall occupancy of the sub detector, estimated to be 3.5\%. There is also an ongoing effort to 
understand if it is possible to benefit from the use of cluster counting in order to improve
the $dE/dx$ resolution by a factor of two~\cite{cluster1,cluster2}.  No experiment has yet managed to utilize this technique,
so the potential use of cluster counting is considered an interesting option for improving 
the DCH performance beyond an acceptable design level.  The DCH will provide fast trigger information
for events with charged tracks in the final state.

  \item{\bf PID}:
A next generation Detector
of Internally Reflected Cherenkov radiation (FDIRC) surrounds the 
DCH tracking volume, and is used to distinguish between different types of charged particle.  The 
fused silica bars of the \babar DIRC will be re-used in \superb, and combined with a segmented
fused silica focusing block system read out by semiconductor sensors.  This next generation
DIRC will have superior particle identification performance and tolerance to backgrounds than the
first generation \babar DIRC.  The reason for replacing the water based focusing system with 
a fused silica one is to produce a radiation hard device that has a much smaller instrumented area
than the \babar DIRC, in order to be able to cope with increased levels of 
backgrounds that are expected at \superb.  There is an option for a forward PID
system, which is motivated by the desire to increase the angular coverage of the PID system 
and there are several technologies under study for a potential device.  
The additional material required to implement such a system would not 
significantly degrade the performance of the calorimeter end-cap.

  \item{\bf EMC}:
The primary purpose of the EMC is to measure the energy deposited by photons and 
electrons that have traversed the inner regions of the detector.
There are two parts to the EMC, a barrel and an end-cap.  \superb will reuse 
the \babar CsI(Tl) crystal calorimeter barrel as the 
crystals themselves are adequate for the rates, and fluence of particles expected
at \superb, however the readout electronics will be updated in order to operate
at a suitable rate.  The end-cap part of the calorimeter needs to be replaced
with a faster and more highly segmented solution than that used for \babar. 
The baseline choice of crystal in the \superb end-cap calorimeter is Cerium 
doped Lutetium Yttrium Orthosilicate (LYSO) which provides a superior performance over all
other materials currently available.  Different options are being considered 
as alternate solutions, such as CsI, PbWO, BGO. All of these crystals have significant 
lower light yield than LYSO, but studies are ongoing to evaluate if they can 
provide acceptable operational performance. 
The amount of material in the barrel of the EMC
is between 16.0 and 17.5 radiation lengths, and will be 17.5 radiation lengths
for the end cap.

  \item{\bf IFR}:
The purpose of the IFR is twofold; to detect and positively identify muons, and to
detect $K_L$ mesons that would have passed through the inner part of the detector and 
typically interact in absorber material between the active layers of this sub-system.
The baseline technology chosen for the high rate environment expected at \superb
is based on wavelength shifting scintillating fibres that are read out
by avalanche pixel photo-diodes operating in Geiger mode.
Studies are ongoing in order to determine the amount of iron required for optimal
muon identification performance at \superb.  It was recognised that there was insufficient
material used at \babar to optimally detect muons, and it is intended that more 
material will be used in the \superb IFR.  The amount of material in the
IFR will be about 5.5 radiation lengths.

  \item{\bf Trigger}: Unlike a hadron experiment, most of the interactions
occurring at an $e^+e^-$ collider like \superb are of interest.  One would only
want to record a small fraction of the Bhabha scattering events that occur,
given the large cross-section for this process, relative to other events of
interest. As a result 
there is a strong constraint on the trigger performance.  The experiment 
must be able to trigger on as many interesting events as possible, in 
the first instance (Level 1), so that a subsequent software filtering system
can be applied to events at the so-called Level 3 trigger stage.  Currently
there is no intention to introduce a Level 2 trigger in \superb. Events
that pass the Level 3 trigger are stored for full event reconstruction 
and offline analysis.  A Level 4 trigger, a higher level software trigger 
used to make decisions to reduce the volume of data permanently stored, may
be implemented if required.  At this time however the intention is to 
implement an open trigger system that records almost all of the interesting
physics events occurring in the detector.  The present expectations of 
the trigger system are to have a Level 1 accept rate of 150kHz, with 
an event size of 75kb, and an anticipated dead time of less than 1\%.
\end{description}

As with any modern particle physics experiment, \superb will produce hundreds of petabytes of data 
during its lifetime, equivalent to an \lhc experiment.  
Thus in order to be able to record and subsequently analyse data in an efficient
way, this experiment is already using the latest GRID technology.  There are two Monte Carlo simulation 
programmes available for SuperB.  The first is based on GEANT4~\cite{GEANT4}, and the second is a third 
generation Fast Simulation (FastSim) with track fitting and other advanced capabilities~\cite{FastSim}.
A more detailed description of the \superb detector can be found in Ref.~\cite{detectorwp}.
Both of these simulations are used in order to compute the projected sensitivities discussed
in the remainder of this document.

%% file: tau.tex
\section{\boldmath{$\tau$ physics}}
\label{tau}

The \sffs will record vast quantities of $\tau$ lepton decays that can be used to 
probe our understanding of a number of areas.  These include searches for 
charged Lepton Flavour Violation, otherwise referred to as LFV 
(Section~\ref{sec:tau:lfv}), \CP violation
(Section~\ref{sec:tau:cpv}), precision measurements of the electric dipole 
moment and $g-2$ of the $\tau$ (Section~\ref{sec:tau:edmetc}), as
well as precision measurements of SM quantities such as \vus\
(Section~\ref{sec:tau:vus}).  The polarised electron beam at
\superb provides a statistical advantage over experiments with 
unpolarised beams.  With polarised electrons, one can reconstruct 
the $\tau$ helicity distribution and thus use information on
the polarised $\tau$ final states to suppress backgrounds.  This 
can be particularly relevant when searching for forbidden or rare 
processes, but is also useful for measurements of the electric dipole
moment and $g-2$.

\subsection{Lepton Flavour Violation}
\label{sec:tau:lfv}

Lepton flavour changing processes were traditionally forbidden in the SM,
however since the discovery of neutrino oscillations, one has had to account
for not only lepton number violation in the neutrino sector, but also the intrinsic
possibility that charged lepton number should also be violated.  If neutrino 
oscillations are the sole source of charged lepton number violation, then it
is unlikely that any experiment would be able to reach the required
sensitivities to probe such effects in the foreseeable future.  
Given that both quark and neutrino flavour numbers are already known to be violated
at a small level, it is reasonable to assume that the corresponding scenario could also
be true in the charged lepton sector. In fact many scenarios of 
physics beyond the SM allow for large enhancements of charged LFV, 
and as in the case of the neutrino sector affecting expectations in the 
charged lepton sector, many models of charged LFV are dependent also on 
the neutrino mixing parameters.  In order to full determine the underlying nature
of any NP affecting the lepton sector one will have to combine results from
both $\tau$ and $\mu$ decay studies with results from the neutrino sector.

In terms of charged LFV, one needs to constrain all three sets of possible transitions between
generations ($2\to 1$, $3\to 1$, and $3\to 2$), in analogy with 
the programme of experiments studying neutrino oscillations currently 
underway. Tests of LFV transitions from the second to first generation can be
performed by searching for the decay $\mu\to e \gamma$, or through searches
for $\mu\to e$ conversion in the presence of nuclear material.
The MEG experiment at the Paul Scherrer Institut (PSI) in Switzerland
is searching for $\mu\to e \gamma$ transitions~\cite{meg}, 
and the most precise constraint on $\mu\to e$ conversion comes
from the Sindrum II experiment~\cite{mu2e:sindrum}, which was also
based at PSI.  Two new $\mu\to e$ conversion experiments, COMET and 
PRISM, are being planned.  
Searches for charged LFV using $\tau$ decays focus 
on transitions from the third to the first and second generations 
of charged leptons.  There are a number of different models of 
NP that generally predict which set of measurements will provide the 
most stringent constraints on charged LFV.  While these models are
a useful guide to follow, one should remember that historically 
in the neutrino sector, oscillation results have often been contrary 
to the most popular scenarios initially proposed,
and secondly it is not simply good enough to
observe LFV in one physical process.  In order to understand 
the structure of NP in a detailed way, one needs to over constrain
couplings related to LFV transitions, and hence to study the
full set of possible lepton flavor transitions.  
\superb is able to contribute to this area by searching for charged LFV 
in the decay of $\tau$ leptons.  
The combination of results from \sffs
with those from MEG and future $\mu\to e$ conversion experiments,
will provide a powerful set of constraints on the sets of charged LFV
transitions.  

In the SM, neglecting the tiny contribution from neutrino mixing, the decay 
of a $\tau$ lepton necessarily results in at least one neutrino in the final 
state via the decay of a $\tau$ into a virtual $W$ boson
and a $\nu_{\tau}$. The virtual $W$ boson subsequently decays into either 
a hadronic state or a lepton-neutrino pair.
A Lepton Flavour Violating (LFV) decay involves the
direct transition of the initial $\tau$ to a final state devoid of a neutrino.
The two most important examples of LFV decays to be studied at \superb are
\begin{eqnarray}
 \tau^\pm \to \ell^\pm \gamma, \nonumber \\
 \tau^\pm \to \ell^\pm \ell^+\ell^-, 
\end{eqnarray}
where $\ell=e, \mu$.  As neutrinos are known to change flavour, it is possible 
for a $\tau$ to decay into one of the above final states, without emitting a 
neutrino.  However the SM expectations for such a LFV branching fraction
is well beyond current experimental reach, for example the branching fraction
for $\tau\to\mu\gamma$ or $\mu\to e\gamma < 10^{-40}$~\cite{PhysRevLett.45.1908},
where current knowledge of neutrino mixing places this limit closer to $\sim 10^{-54}$.
In many popular scenarios of new physics these LFV branching fractions can be
enhanced to the level of 
$10^{-9}$~\cite{Allanach:2002nj,Masiero:2002jn,Hisano:2003bd,Ciuchini:2003rg,Arganda:2005ji,Paradisi:2005tk,Antusch:2006vw,Dreiner:2006gu,Parry:2007fe,Ciuchini:2007ha,Arganda:2008jj,Calibbi:2008qt,Raidal:2008jk,Calibbi:2009ja,Herrero:2009tm}, 
for example by introduction 
loop contributions containing sparticles, where experimental
bounds limit the parameter space of the models.

The $\ell \gamma$ and $3\ell$ decays are golden channels for \superb and  
the experimental reach for these
modes is given in Table~\ref{tbl:tau::lfv},
where the results from \babar and \belle come from Refs.~\cite{Aubert:2009tk,Lees:2010ez,Hayasaka:2007vc,Hayasaka:2010np}.  In 75\invab of data collected at
the \FourS \superb expects to record $70\times 10^9$ $\tau$ leptons pairs.
A further $\sim 1.2\times 10^9$ $\tau$ pairs should be accumulated 
during the 500\invfb run at charm threshold.
As one can see, the experimental sensitivity of these golden
modes will be able to constrain NP models between one and two
orders of magnitude better than current bounds, and
interpretation of limits from data in the context of such models 
is discussed in Section~\ref{sec:tau:lfv:np}.
The expected limits achievable by \belletwo are slightly
worse than those indicated for \superb for two reasons,
firstly \belletwo will accumulate slightly less data than
\superb, and secondly the electron beam will be longitudinally polarised in 
\superb, introducing an additional kinematic variable: the polarisation of the $\tau$.
The utilisation of this information when analysing data will 
provide \superb with an additional variable to suppress background
relative to the signal that is not available at other proposed or
existing experiments.

It should also be noted that the relative difference between the 
existing \B factory limits for $\ell \gamma$ and $3\ell$, and expectations
for \superb are similar for many other LFV $\tau$ decay measurements
that \superb will be able to perform.

\begin{table}[!h]
\caption{Experimental sensitivities of Lepton Flavour Violation searches at \superb in comparison with existing
limits from \babar and \belle.}
\label{tbl:tau::lfv}
\begin{center}
\begin{tabular}{lccc}\hline
Mode                              & \babar ($\times 10^{-8}$) & \belle ($\times 10^{-8}$)& \superb ($\times 10^{-8}$) \\ \hline
$\tau^\pm \to e^\pm \gamma$       & 3.3 & 12 & 0.3\\
$\tau^\pm \to \mu^\pm \gamma$     & 4.4 & 4.5 & 0.2\\
$\tau^\pm \to \mu^\pm \mu^+\mu^-$ & 3.3 & 2.1 & 0.08\\
$\tau^\pm \to e^\pm e^+e^-$       & 2.9 & 2.7 & 0.02\\
 \hline
\end{tabular}
\end{center}
\end{table}

The \lhcb experiment may ultimately be able to reach a 
similar sensitivity to the current limits from 
\babar and \belle in the $\tau\to 3\mu$ channel, but 
will not be competitive with other channels.  There is 
also a possible upgrade to \lhcb 
that is under consideration~\cite{lhcbupgradeloi}, however it is clear that even in 
the best case scenario any \lhcb upgrade will be an order 
of magnitude less sensitive than the searches from the \sffs,
based on the results presented in~\cite{lhcbtauphysics}.
The relatively poor performance from \lhcb is a direct result
of the hadronic environment at \lhc, where there are no primary 
vertex constraints available to kinematically separate signal 
from background as cleanly as one can at a \sff.

\subsubsection{New physics scenarios and LFV}
\label{sec:tau:lfv:np}
$\,$

\begin{figure}[!ht]
\begin{center}
  \includegraphics[width=10cm]%
        {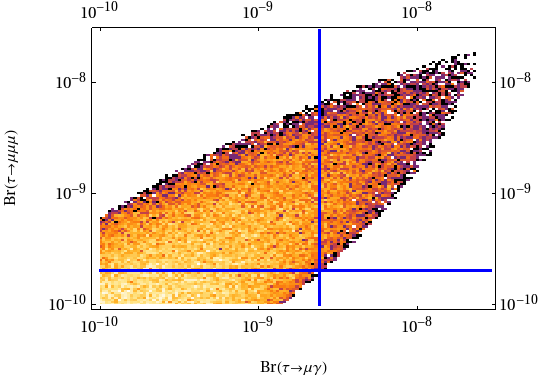}
  \caption{The branching fraction for $\tau\to \mu\gamma$ vs. $\tau\to 3\mu$ in Little Higgs model with T parity,
shown with the expected sensitivity reach from \superb (solid lines), from Ref.~\cite{Blanke:2009am}.}
  \label{bevan:fig:taulht}
\end{center}
\end{figure}

Any evidence found at existing or planned experiments for 
charged LFV would be a clear sign of physics beyond the 
SM, which in itself would be a significant milestone in particle physics.
However one can go beyond this binary test and try to 
elucidate the dynamics of possible new physics scenarios, both in the presence
and absence of a signal.
Given that the interpretation of LFV is highly model dependent, one needs to
identify realistic benchmark models to test using data, and a number of observables that
can be used to distinguish between the benchmarks.  A lot of work has been done in this
area, which include CMSSM and NUHM SUSY as specific variants of a more
generalised SUSY model, as well as little Higgs models (LTH) and SUSY
GUT models~\cite{Allanach:2002nj,Masiero:2002jn,Hisano:2003bd,Ciuchini:2003rg,Arganda:2005ji,Paradisi:2005tk,Antusch:2006vw,Dreiner:2006gu,Parry:2007fe,Ciuchini:2007ha,Arganda:2008jj,Calibbi:2008qt,Raidal:2008jk,Calibbi:2009ja,Herrero:2009tm}.
By piecing together different experimental observations
one can distinguish between different types of model.  For example the channel $\tau^\pm \to \ell^\pm \gamma$
can be significantly enhanced in SUSY based models relative to $\ell^\pm\ell^\mp\ell^\pm$, whereas 
$\tau^\pm\to\ell^\pm\ell^\mp\ell^\pm$ can have a corresponding enhancement in 
LTH models relative to $\ell^\pm \gamma$. Hence a measurement of the ratio of the rates of these two 
channels could be used to distinguish between these sets of models.
A recent analysis of the correlation between the $\mu\gamma$ and $3\mu$ branching fractions 
(shown in Figure~\ref{bevan:fig:taulht}) for Little Higgs models with T parity can be found in Ref.~\cite{Blanke:2009am}.  
\superb will be able to exclude all but the bottom left quadrant of the phase space shown
in the Figure.

Two examples of how results from \superb and other 
experiments can be combined are given in the following.  If one considers the $SU(5)\times T^\prime$ 
model of Ref.~\cite{Chen:2007afa}, there is a definite
prediction that lepton flavour violation in the MSW sector arises from charged
leptons and that $\nu$'s only mix.  Other models can be built where this is 
not the case.  This model also has well defined 
predictions for \CP violation in \B and \D decays.
Hence it is important to make as complete a set of measurements
as possible in the charged lepton and quark sectors, and consider the 
relationship between such measurements and the $\nu$ sector to verify or 
refute this prediction. 
Similarly in the model of Antusch et al.~\cite{Antusch:2006vw}, 
the rates of $\tau^\pm \to \ell^\pm \gamma$,
$\mu\to e \gamma$, and the neutrino mixing parameter $\sin \theta_{13}$
are correlated, once again highlighting the need for a global approach
to interpreting results.  The recent $\nu_e$ appearance result
from the T2K experiment~\cite{Abe:2011sj} is an important step forward in neutrino physics, and future
updates of that result will be of interest when considering possible new physics
scenarios in the context of charged lepton flavour violation.
If $\theta_{13}$ turns out to be large as currently suggested by data, 
then according to this model \superb should not see $\tau \to \mu \gamma$.  The corollary of this is that if one were to 
observe a large signal for this channel, then this model would be 
ruled out.

\subsection{CP violation}
\label{sec:tau:cpv}

In the SM $\tau$ decays proceed via a single amplitude, and hence there is no 
\CP violation. 
The exceptions to this rule are decays to final
states including kaons, where the well known level of \CP violation in kaon decay may be 
manifest when reconstructing the final state.
Any significant deviation found in a measurement of a \CP asymmetry in the decay of a $\tau$ lepton
would be an unequivocal sign of NP.  Models of NP that can naturally manifest \CP 
violation in $\tau$ decays include multi-Higgs models~\cite{Kuhn:1996dv}, which could modify
the angular distributions of decaying $\tau$ leptons relative to SM expectations.
This area has been largely unexplored, and so far results are only available
for $\tau^\pm\to\KS\pi^\pm\nu$ decays~\cite{Bischofberger:2011pw,BaBar:2011mj}, where
the SM expectation of the \CP asymmetry is $(0.33\pm 0.01)\%$~\cite{Bigi:2005ts}.  The \babar result is
$(-0.45\pm 0.24\pm 0.11)\%$, and \belle report results as a function of the $\KS\pi$ mass
that are compatible with zero.

The extraction of \CP asymmetry parameters is complicated by 
having to understand matter-antimatter effects in the detector.  The reason for this is that
both matter and
antimatter have slightly different cross sections for interaction within the
detector, which is constructed entirely of matter.  A detailed understanding of 
this difference is straightforward and can be modeled in simulation.  Another 
way to understand the magnitude of such a matter-antimatter asymmetry effect 
is to use a calibration sample of data.  The advantage of this is that one
can remove reliance of these important measurements on any simulation, and 
this route has been taken with existing searches for \CP violation.
It will be possible for \superb to significantly improve upon the 
precision of \CP violation searches in $\tau$ decays.  In the case of the 
$\tau^\pm\to\KS\pi^\pm\nu$ decay, where there is an intrinsic \CP asymmetry
resulting from kaons in the final state, \superb should be able to reduce
current upper limits to the level where this SM background effect becomes
measurable.
One can use the channel $\tau^\pm\to\pi^+\pi^-\pi^\pm\nu$ 
in a region where the signal \KS has been vetoed to control systematic uncertainties
arising from any detector asymmetry for the $\KS\pi^\pm\nu$ channel.  
Indeed the existing experiments are approaching that sensitivity now, 
and the \babar result of almost $3\sigma$ deviation from the SM 
motivates higher statistics searches. The ultimate precision achievable for 
\CP asymmetry measurements in $\tau$ decays needs to be evaluated, both in terms
of integrated rate measurements and in terms of angular distributions.

\subsection{Electric dipole moment and $g-2$ of the $\tau$ lepton}
\label{sec:tau:edmetc}

The electric dipole moments (EDMs) of charged leptons $d_\ell$ are sensitive to different 
models of new physics including MSSM, generic SUSY, and multi-Higgs extensions of the 
SM.  As is the case 
with lepton flavor violation measurements it is necessary to measure all three of the EDMs
in order to understand which model or underlying mechanism may be at play in physics
beyond the SM.  For example the different values of $d_\ell$ scale with lepton mass
in the case of MSSM, where any new \CP phases are independent of flavour~\cite{Calibbi:2008qt}, 
whereas more general models of SUSY could produce large effects for the $d_\tau$,
and small effects for both $d_e$ and $d_\mu$~\cite{Pilaftsis:1999qt}.  The values of $d_\ell$ in 
multi-Higgs models scale as $m_\ell^3$~\cite{Barger:1996jc}.

One can measure the $\tau$ EDM using an angular asymmetry in $\epem\to \tau^+\tau^-$ 
transitions. The current limit on the $|\tau|$ EDM is $\leq 5 \times10^{-17}e$cm and comes 
from the \belle Collaboration~\cite{Inami:2002ah}, using a data sample of 29.5\invfb.
The anticipated reach for \superb using 75\invab of data
is $d_\tau \leq 17 - 34\times 10^{-20} e$cm without a polarised electron beam.  One can
improve the sensitivity by almost a factor of two beyond this with 80\% electron beam polarisation.

The anomalous magnetic moment measured for the muon is not in good agreement with 
the SM. The difference between the SM and experimental measurement is 
$\Delta a_\mu = \Delta a_\mu^{\mathrm{expt}} - \Delta a_\mu^{\mathrm{SM}} = (3\pm 1)\times 10^{-9}$.
A measurement of the anomalous magnetic moment for the $\tau$ would enable us
to understand if $\Delta a_\mu$ is the result of NP, or simply a statistical fluctuation 
in data.  In NP scenarios one expects $\Delta a_{\mu, \tau}$ to scale with the lepton mass
squared, and so one would anticipate $\Delta a_\tau \sim 10^{-6}$ if the muon signal was
an indication of NP.  In fact $\Delta a_{\tau}$ can be as large as $10^{-5}$ in 
some NP scenarios.  With a polarised electron beam at \superb one will be able to 
measure $\Delta a_\mu$ to a statistical precision of $2.4\times 10^{-6}$ from 
$\epem\to \tau^+\tau^-$ transitions~\cite{physicswp}.

\subsection{Measurement of \Vus}
\label{sec:tau:vus}

Up until recently knowledge of \Vus\ has been dominated by results from
studies of kaon decays.  This approach is
limited by theoretical uncertainty, and \Vus\ has been measured to 
$\sim 0.8\%$~\cite{Nakamura:2010zzi}, where the experimental contribution
is $\sim 0.2\%$.  The opposite scenario is encountered in $\tau$ decays, where the 
theoretical uncertainty is relatively small, and the experimental uncertainty 
dominates~\cite{Gamiz:2007qs}. It will be possible to produce a more
precise constraint on this SM parameter using $\tau^\pm\to K^\pm\nu$ decays
at \superb and the potential for this is currently under study.  The
current determination of \Vus\ from kaon and $\tau$ decays is~\cite{Nakamura:2010zzi,Gamiz:2007qs}
\begin{eqnarray}
\vus (K) &=& 0.2255 \pm 0.0004 \expt \pm 0.0019 \theo, \\
\vus (\tau) &=& 0.2165 \pm 0.0026 \expt \pm 0.0005 \theo.
\end{eqnarray}

A precision measurement of \Vus feeds into testing a unitarity constraint on the 
CKM mechanism.  In fact this element also plays a role in the charm $cu$ triangle,
that has yet to be tested directly (see Section~\ref{dphysics}).  While 
\Vus is not the limiting constraint on this test of CKM unitarity, it is 
apparent that an improved precision on this quantity will be desirable 
on the time scale of the \sffs.

%% file: bphysics.tex
\section{\boldmath{B physics}}
\label{bphysics}

\B physics at \superb is divided into the study of the decays of $B_d^0$ and $B_u^\pm$ mesons
produced via $\epem\to\FourS\to\B\Bb$, and the study of $B_d^0$, $B_u^\pm$ and $B_s^0$ mesons as well 
as excited states via $\epem\to\FiveS\to\B^{(*)}\Bb^{(*)}$.  Selected highlights of 
\FourS and \FiveS programmes are discussed in Sections~\ref{bphysics:4S} and \ref{bphysics:5S}, respectively.

\subsection{B physics at the $Y(4S)$}
\label{bphysics:4S}

One might feel justified in asking, why bother with a second generation Super Flavour Factory
programme of \B physics given the successes of \babar and \belle, and the potential of the CERN 
based \lhcb experiment.  It is true that the CKM mechanism has been verified at the 10\% level,
resulting the Kobayashi and Maskawa being awarded the Nobel Prize for physics in 2008 for 
their innovative work on a three generation quark mixing matrix to introduce
\CP violation into the SM.  However one should not overlook the fact that new possibilities present
themselves with one hundred times more data than existing experiments.  \superb will
accumulate $75\times 10^{9}$ neutral and charged $B$ mesons in a data sample of 75\invab, 
compared with $0.4-1.0\times 10^{9}$ accumulated by \babar and \belle. With this increase
in data it will be possible to perform a \% level test of the CKM mechanism, and substantially 
improve a number of constraints on NP.  The \lhcb experiment will make a number of important
measurements in the coming years, however as previously mentioned there are 
complementary advantages (and disadvantages) for experiments operating 
in \epem environments versus hadronic ones that mean \lhcb is complementary
to \superb, as opposed to a natural competitor.  The true strength of these two experimental
programmes comes when one combines the total set of observables that they will be able to measure.
Here we concentrate on the contributions that \superb will make to understanding new physics.
One should also note that there are a number of 
issues that have recently been raised with regard to existing measurements of \B decays
that need to be explored in greater detail, as these may already be indications of problems
with the SM:
\begin{description}
  \item {\bf \boldmath $\sin 2\beta$:} The measured value of this quantity from $B$ meson decays to final
    states including a neutral kaon and charmonium ($c\overline{c}$) is $3.2\sigma$ from the 
    SM preferred value as highlighted in Ref.~\cite{Lunghi:2011xy}.
  \item {\bf \boldmath \Vub and \Vcb:} 
    There are disagreements between inclusive and exclusive results for \Vub and \Vcb.  These
    are only resolvable with a more thorough analysis provided by increased data samples, and improvements
    in theoretical understanding.
  \item {\boldmath \CPT}: The \CPT asymmetry measurement as a function of sidereal time 
    made by \babar is $2.8\sigma$ from SM expectations and could indicate NP~\cite{Aubert:2007bp}.
  \item {\boldmath $A_{\mathrm{SL}}$:} 
    Semi-leptonic asymmetry measurement made recently by the D0 experiment is $3.9\sigma$ from
    SM expectations~\cite{Abazov:2011yk}.
  \item {\boldmath $B_s\to \mu^+\mu^-$:}
    The Tevatron reports an excess of $B_s\to \mu^+\mu^-$ well above the expected SM branching 
    fraction~\cite{Collaboration:2011fi}, however data from CMS  and \lhcb presented at EPS 2011~\cite{epsmumu}
    suggest that the Tevatron data is probably the result of a background fluctuation.
  \item {\bf New physics energy scale:} $\Lambda_{NP}$ is widely believed to be $\sim 1\mathrm{TeV}$, in order to resolve the so-called hierarchy problem.  This is a regime that is currently 
    being probed by the \lhc.  Flavour observables indicate that $\Lambda_{NP}$ may be significantly
    higher than the TeV scale, where the scale indicated is model dependent.  Recent results from the \lhc 
    are placing considerable constraints on the NP parameter space, and it is looking increasingly
    likely that any new physics discovery may not be just around the corner as was once believed 
    to be the case.
    If this is indeed the case, then one needs to either (i) build an energy upgrade to the
    \lhc, (ii) a sufficiently high energy \epem linear collier, or (iii) use indirect constraints to constrain
    new physics.  Neither of the first two options will be easy or quick to achieve.
    The third option is based on the role \belletwo,
    \superb, and other flavour physics 
    experiments can play in placing model dependent constraints on high energy physics.  However,
    if NP is discovered at the \lhc, a similar indirect methodology can probe the mixing couplings of
    particles related to the NP sector.  More details of this route can be found in Refs.~\cite{physicswp,Aushev:2010bq}.
\end{description}
These discrepancies may ultimately turn out to be statistical fluctuations, or in some cases the result of some
mis-understanding in the SM description of the observable, however it is clear
that there are a number of unresolved issues in $B$ physics that need to be pursued, and some of these
can only be addressed using experiments at an $\epem$ collider.

The $B_{u,d}$ physics programme at \superb is large, and for brevity, the following discussion is confined 
to the study of rare \B decays (Section~\ref{bphysics:4S:rare}), precision angle and sides measurements 
(Section~\ref{bphysics:4S:angles}) of the $bd$ unitarity triangle.  
The forthcoming Physics of the B Factories book currently
in preparation~\cite{PBF} will describe many of the additional measurements of interest.

\subsubsection{Rare \B decays}
\label{bphysics:4S:rare}
$\,$

There are a number of rare \B decays of interest at \superb.  Most of the interesting final
states contain neutral particles such as photons that are best studied in an \epem\ environment
and final states with neutrinos, which can only be studied in an \epem environment.  
Only a selection of these decays are discussed in the following: 
$B\to K^{(*)}\nu\overline{\nu}$, $B\to \ell\nu$, $b\to s\ell\ell$, and $b\to (s,d)\gamma$
(where $\ell=e, \mu, \tau$) and more information can be found in Ref~\cite{physicswp}.

{\boldmath {$B\to K^{(*)}\nu\overline{\nu}$}:}
Decays to final states with $\nu\overline{\nu}$ allow one to study $Z$ penguin transitions.  In the SM
the $K^*$ ($K^+$) channel has a branching fraction of $6.8\times 10^{-6}$ ($3.6\times 10^{-6}$)~\cite{Altmannshofer:2009ma,Bartsch:2009qp}.
The cleanest theoretical observables are the inclusive 
branching fraction measurement $B\to X_s \nu\overline{\nu}$ and the fraction of longitudinally $f_{\mathrm{L}}$
polarised events in $B\to K^{*}\nu\overline{\nu}$ decays, however the former is a challenging 
measurement, while measurement of the latter first requires one to observe the decay mode.
The branching fractions and $f_{\mathrm{L}}$ 
in the $B\to K^{*}\nu\overline{\nu}$
mode are sensitive to NP.  Large effects could result from models with right handed currents, $Z^\prime$ bosons,
and models with light scalar particles~\cite{Adhikari:1994wh,Bird:2004ts,Davoudiasl:2009xz,Dreiner:2009er,Hiller:2004ii}.  
In contrast, only small effects are found in models of
minimal flavor violation such as MSSM~\cite{Altmannshofer:2009ma,Yamada:2007me}. These decays can be parameterised in terms of left and 
right handed Wilson coefficients, $C^\nu_{\mathrm{L, R}}$ via $\epsilon$ and $\eta$, where
\begin{eqnarray}
\epsilon = \frac{\sqrt{|C^\nu_{\mathrm{L}}|^2 + |C^\nu_{\mathrm{R}}|^2}}{|C^\nu_{\mathrm{L}}|^{\mathrm{SM}}}\,\,\, \mathrm{    and    }
\,\,\, \eta = -\frac{Re(C^\nu_{\mathrm{L}} C^\nu_{\mathrm{R}})}{|C^\nu_{\mathrm{L}}|^2 + |C^\nu_{\mathrm{R}}|^2}.
\end{eqnarray}
The branching fractions and $f_{\mathrm{L}}$, relative to their SM expectations, can be 
parameterised in terms of $\epsilon$ and $\eta$, where $(\epsilon, \eta)_{\mathrm{SM}} = (1,0)$.

\begin{figure}[!ht]
\begin{center}
  \includegraphics[width=10cm]%
        {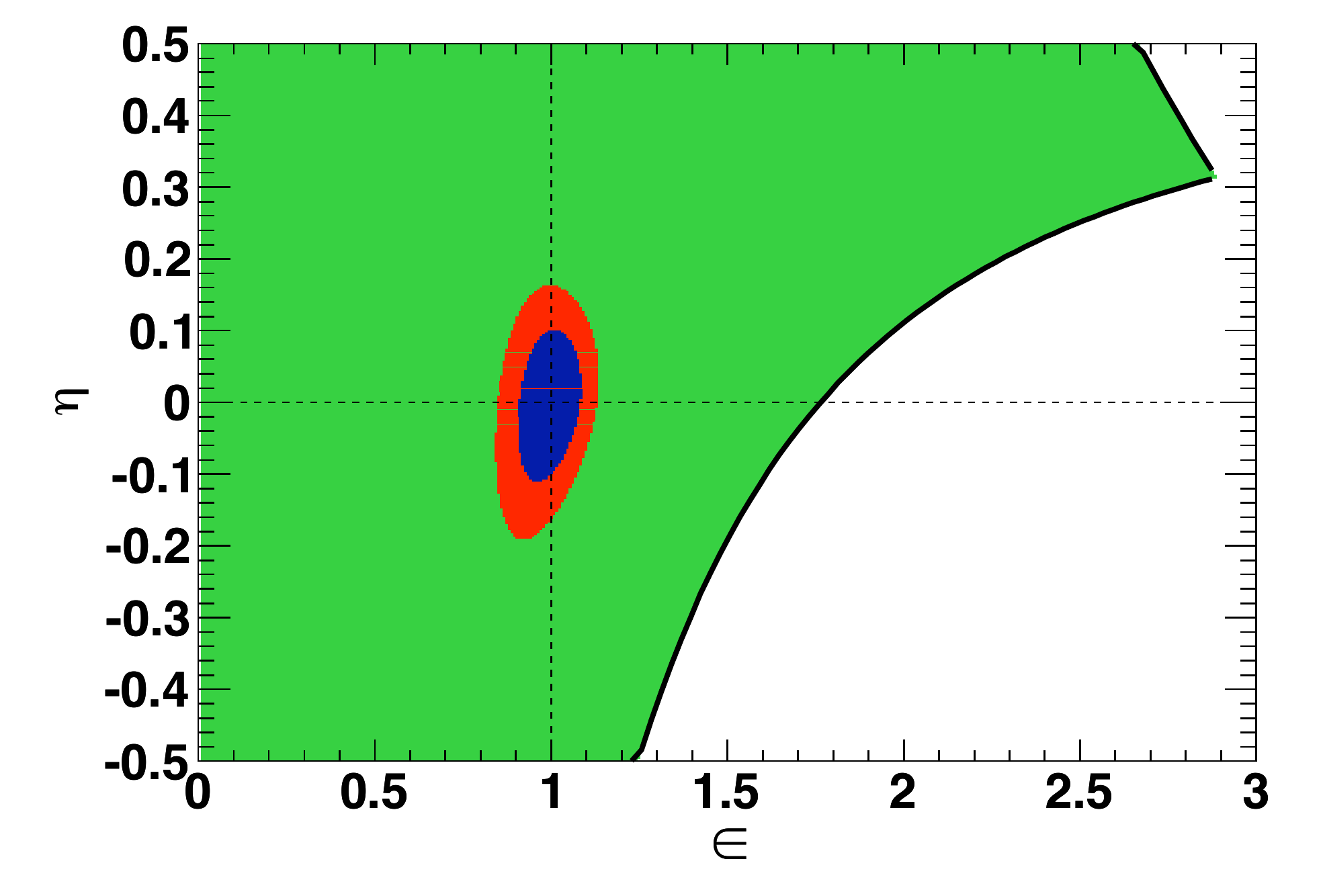}
  \caption{The constraint on $\epsilon$ and $\eta$ expected using exclusive branching fraction and $f_{\mathrm{L}}$
  measurements made with data sample of 75\invab at \superb (from Ref.~\cite{Meadows:2011bk}).
  The central two contours represent the 68\% and 95\% confidence level (C.L.) constraint 
  obtained at \superb, while the light (green) contour indicates
  the existing constraint obtained using limits on the $B\to K^{(*)}\nu\overline{\nu}$ modes.}
  \label{fig:knunubar}
\end{center}
\end{figure}

With 75\invab of data at \superb one should be able to make measurements of the branching fractions 
of the exclusive modes at the 16-20\% level.  Figure~\ref{fig:knunubar} shows the constraint
expected on the $(\epsilon, \eta)$ plane using exclusive branching fraction and $f_{\mathrm{L}}$ 
measurements at \superb.
In order to achieve this level of precision one has 
to have a good hermiticity of the detector, to limit the level 
of background.  Given that only a rudimentary measurement 
of $f_{\mathrm{L}}$ will have been made with this data sample, one could envisage the desire to 
perform a high precision study of these decays with data samples of hundreds of \invab in 
the longer term.  Such a measurement would greatly improve the constraint on $\eta$.

\begin{figure}[!ht]
\begin{center}
  \includegraphics[width=10cm]%
        {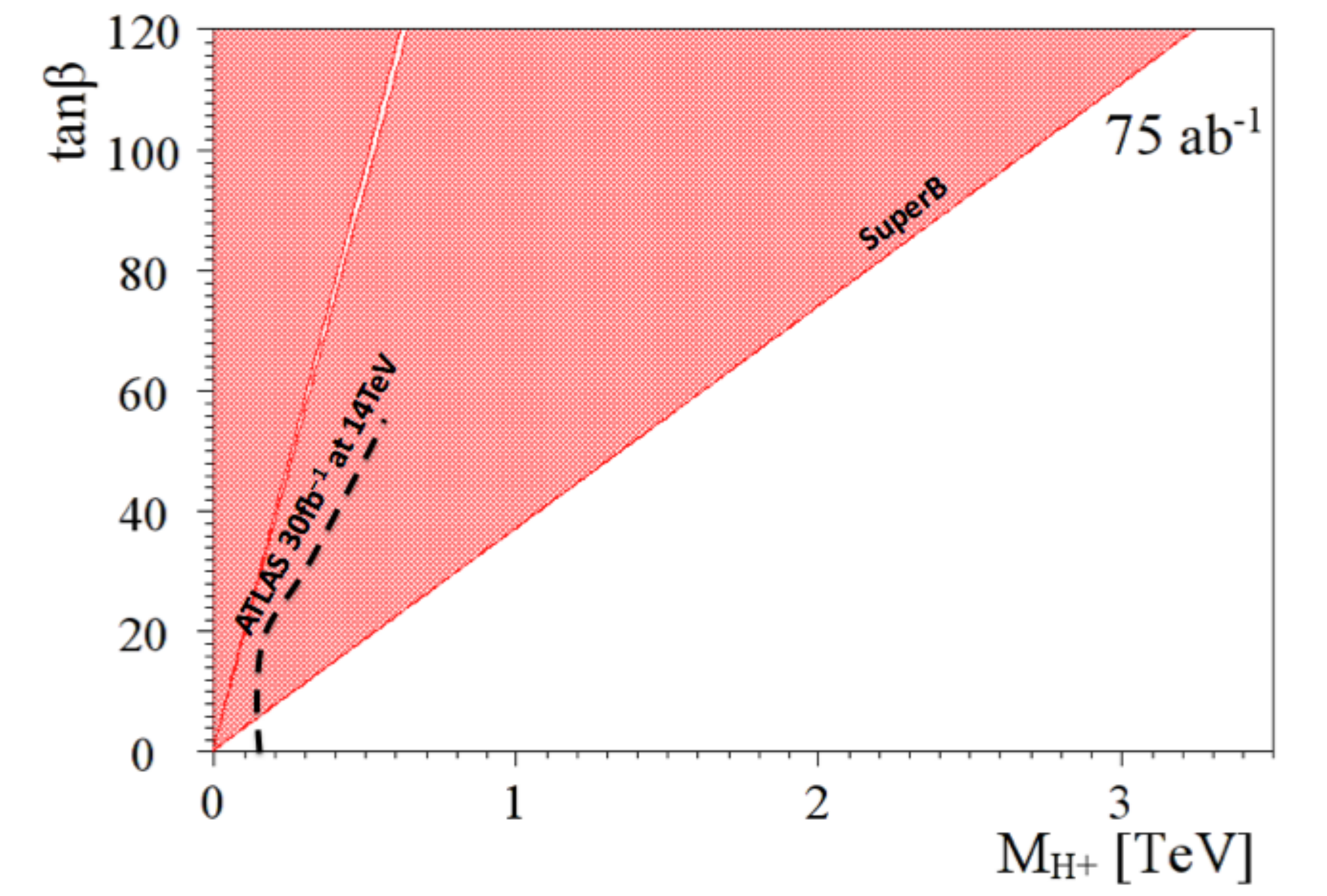}
  \caption{The constraint on the $m_{H^+}-\tan\beta$ plane from $b\to \ell\nu$ branching fraction 
    measurements at \superb, compared with the expectations of direct searches at the LHC.  The shaded
    region will be excluded by $b\to \ell\nu$ decays.  Measurements from $b\to s\gamma$ already exclude
    values of $m_{H^+}<295\gevcc$}
  \label{bevan:fig:ellnu}
\end{center}
\end{figure}

{\boldmath {$B\to \ell\nu$}:}
In the SM the branching ratio of the set of leptonic decay modes $b\to \ell\nu$ is related 
to \Vub, and can be computed using Lattice input on the parameter $f_B$.  Hence this channel
can be combined with other determinations of CKM parameters in order to test the SM.  If one
considers NP scenarios with Higgs multiplets, then one can replace the $W$ boson in the SM
amplitude for this decay with a charged Higgs particle.  The modification to the expected 
rate for this decay depends on both the charged Higgs mass $m_{H^+}$, and on the ratio of 
Higgs vacuum expectation values, $\tan\beta$.  In this scenario it is possible to use a branching 
fraction measurement to indirectly constrain the $m_{H^+}-\tan\beta$ plane.  For a two Higgs
doublet model (2HDM) the branching fraction can be modified by a scale factor $r_H$ relative
to the SM rate, where~\cite{PhysRevD.48.2342}
\begin{eqnarray}
r_H = \left( 1 - \frac{m_B^2}{m_H^2}\tan^2\beta\right).
\end{eqnarray}
The corresponding constraint on the $m_{H^+}-\tan\beta$ plane resulting from measurements
of $b\to \ell\nu$ decays expected from \superb 
is shown in Fig.~\ref{bevan:fig:ellnu} (taken from Ref.~\cite{Meadows:2011bk}).
The expectations of direct searches using 14TeV collision data at the LHC is also
shown on this plot for the ATLAS experiment~\cite{Aad:2009wy}.  While there is a region at
low values of $\tan\beta$ that will not be excluded using $b\to \ell\nu$ decays, one should
remember that existing constraints from measurements of the \CP asymmetry in $b\to s\gamma$ 
events already excludes charged Higgs particles with masses less than 295\gevcc.
More recently LHC direct searches have increasingly ruled out the low $\tan\beta$ scenario.
It is worth noting that the constraints will be dominated by $B\to \tau\nu$ at low
luminosity, however at some point the branching fraction measurement of that mode will become
systematically limited.  As a result the high luminosity constraints will be dominated by
the contribution from $B\to \mu\nu$.

If one considers more complicated extensions, such as SUSY variants, then the bounds on the 
$m_{H^+}-\tan\beta$ plane do change, as does the functional dependence of $r_H$ on $m_{H^+}$ and $\tan\beta$.
However the correction arising from the addition of SUSY particles does not alter the 
conclusion drawn that the indirect constraints on searches for charged Higgs particles
from $b\to \ell\nu$ can exclude a larger parameter space than direct searches from the LHC.
While it is possible to use measurements of rare kaon decays in a similar way, additional
model dependence has to be introduced in order to interpret kaon bounds on $m_{H^+}$ and 
$\tan\beta$, hence the $b\to \ell\nu$ bounds are both more general and more rigorous than the
kaon ones.

{\boldmath {$b\to s\ell\ell$}:}  Existing measurements from \babar and \belle on the forward
backward asymmetry in these decays, while consistent with the SM, are more compatible with
possible NP scenarios.  The NP phenomenology that is possible with these decays is extremely
rich, and beyond the scope of this review, however Ref.~\cite{physicswp} discusses many of
the relevant issues. Both inclusive and exclusive decays can be measured at 
\superb, in both $e$ and $\mu$ final states.  The advantage of being able to perform this 
full set of measurements is that one can constrain all NP sensitive observables. The
set of NP sensitive observables includes forward-backward and isospin asymmetries as well
as ratios of the different leptonic final states.  Recently a number of additional
asymmetries have been added to the list, for example see~\cite{Egede:2008uy}.  The theoretical issues associated
with interpretation of inclusive and exclusive measurements are different, so if a deviation
from the SM were to be found one would want to confirm that the two types of measurement
(inclusive and exclusive) both exhibited this behavior in order to identify the underlying cause.
It is expected that \superb will 
collect $10,000-15,000$ $\B\to K^*\ell^+\ell^-$ events. \lhcb has recently started producing results
on the di-muon channel, and expects to accumulate 8,000 events in a data sample of
5\invfb.  The \epem mode is more challenging in a hadronic environment, and \superb is
expected to accumulate twenty times the number of $K^*e^+e^-$ events than \lhcb. 

In addition to these exclusive measurements, \sffs will be able to perform precision
measurements of inclusive modes, where the attainable precision is under investigation.
The inclusive modes are of interest as one can use the measured branching fraction to 
constrain the NP energy scale in MSSM with mass insertions (Section~\ref{interplay:NPlagrangian:massinsertions}). 
For example if one assumes
that the squark and gluino masses are the same, then by combining inclusive measurements
of $b\to s\ell^+\ell^-$ with inclusive measurements of $b\to s\gamma$ (see below), 
one is able to measure both the real and imaginary part of the mass insertion 
parameter $(\delta^d_{23})_{LR}$, a coupling of $2^{nd}$ to $3^{rd}$ 
left-right squark transitions.  This parameter in turn is related to $\Lambda_{NP}$.

In addition to studying the opposite charge $b\to s\ell\ell$ decays, one can search 
for same sign lepton events, so $b\to s\ell^\pm\ell^\pm$, which would be manifest
through transitions involving Majorana neutrinos.  The search potential for 
such a measurement is greater in an \epem environment compared to a hadronic one, 
as there are smaller backgrounds and a complete set of leptonic final states 
can be studied.  In order to constrain couplings for each of the hypothetical 
Majorana neutrino generations in this scenario, one needs to measure all of 
these different final states.

{\boldmath {$b\to (s,d)\gamma$}:}

The inclusive branching fractions, and \CP asymmetries, of $B$ mesons decaying into $X_{s,d}\gamma$ can be used
to constrain new physics scenarios.  Currently one of the most precise limits on the mass of a charged
Higgs particle in a 2HDM comes from $B\to X_{s}\gamma$, where $m_{H^+} > 295 \gevcc$ at 95\% 
C.L.~\cite{Misiak:2006zs}.
This is the most stringent constraint available on $m_{H^+}$ for low values of $\tan\beta$.
The current constraint obtained from this channel, when combined with existing results from
$B\to \ell\nu$, is able to exclude the possibility of finding a charged Higgs particle at the 
LHC for at least the next few years.  The inclusive
branching fraction can also be used to constrain the compactification scale $R$ in minimal models of universal 
extra dimension scenarios.  The current data give $1/R > 600 GeV$ at 95\% C.L~\cite{Haisch:2007vb}.  Constraints
from $X_{d}\gamma$ complement the information obtained from $X_{s}\gamma$, and for example if one combines
information on the direct \CP asymmetries measured in these two inclusive decays it is possible to determine
NP scenarios based on Minimal Flavour Violation (MFV) from more generic models~\cite{physicswp}.

Experimentally one will be able to measure the inclusive $X_{s}\gamma$ branching fraction to a precision of
about 3\% with 75\invab of data at \superb.  The corresponding precision on the direct \CP asymmetry 
is expected to be $\sim 0.02$.  It is also worth noting that the related channel $B\to \KS\piz\gamma$
can also be used as a null test of the SM, where \superb will reach a precision of 0.03 on the time
dependent \CP asymmetry parameter $S$ (see Section~\ref{bphysics:4S:angles}).  
If one observes a large time-dependent \CP asymmetry
in this decay, this would be a clear sign of NP.    This radiative mode is sensitive to right handed currents
and so complements studies of $B\to K^{(*)}\nu\overline{\nu}$ discussed previously.

$\,$
\subsubsection{Precision CKM: Angles and Sides of the Unitarity Triangle}
\label{bphysics:4S:angles}
$\,$

\begin{figure}[!ht]
\begin{center}
  \includegraphics[width=10cm]%
        {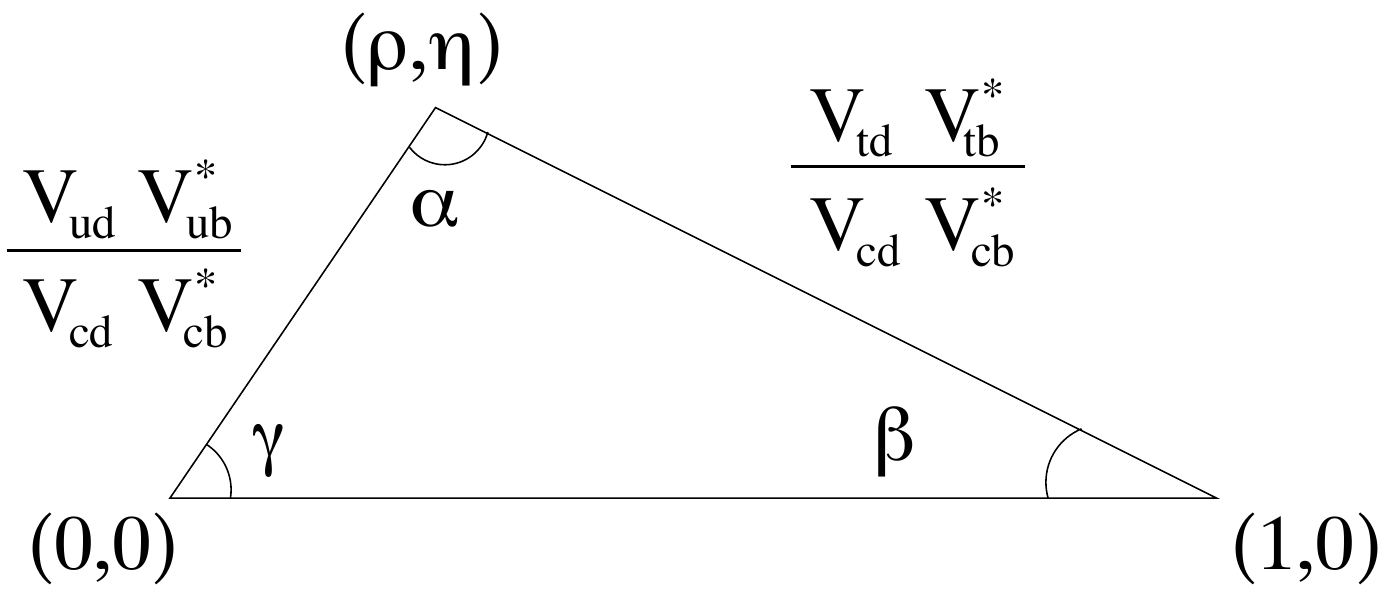}
  \caption{The $bd$ unitarity triangle related to decays of $B_{u,d}$ mesons.}
  \label{bevan:fig:ut}
\end{center}
\end{figure}

Unitarity of the CKM matrix given by Eq.~(\ref{bevan:eq:ckm}) leads to six 
triangles that can be represented in a complex plane.  One of these is related 
to $B_{u,d}$ transitions and can be studied in great detail at a \sff.  This
relation is generally known as the ``unitarity triangle'' and is given by
\begin{eqnarray}
\vud\vub^* + \vcd\vcb^* + \vtd\vtb^* = 0\label{bevan:eq:ut}.
\end{eqnarray}
Here we refer to this triangle as the $bd$ unitarity triangle to avoid possible confusion 
with the $cu$ triangle discussed in Section~\ref{dphysics}.
The triangle itself is shown in Figure~\ref{bevan:fig:ut}, where the base is normalised
to unity, so that any two measurements of the triangle may be used to constrain it
completely.

The angles of the $bd$ unitarity triangle $\alpha$, $\beta$, and $\gamma$ are given by
\begin{eqnarray}
\alpha &=& \arg\left[-\vtd \vtb^*/\vud\vub^*\right], \label{eq:alpha}\\
\beta  &=& \arg\left[ -\vcd\vcb^* / \vtd\vtb^*\right], \label{eq:beta}\\
\gamma &=& \arg\left[ -\vud\vub^* / \vcd\vcb^* \right], \label{eq:gamma}
\end{eqnarray}
and have been measured
with precisions of $6.1^\circ$, $0.8^\circ$, and $11^\circ$ by the 
\B Factories\footnote{One can find an alternate notation in the literature, where 
$(\alpha, \beta, \gamma)=(\phi_2, \phi_1, \phi_3)$, for example in results reported
by the \belle experiment.}~\cite{Bona:2005vz,hfag}.  In addition to precision 
tests of the angles of the $bd$ unitarity triangle, there are measurements of the
sides, where the limiting factors are knowledge of the semi-leptonic
decays $b\to u\ell\nu$ and $b\to c\ell \nu$ which are related to \Vub\ and \Vcb.  
As things currently stand, there is an experimental 
controversy between inclusive and exclusive measurements of these quantities, and
there is a tension between the measurements of $\sin 2\beta$, \Vub, and 
the branching fraction of $\B \to \tau \nu$.
If one artificially moves any one of these parameters to try and mitigate the 
discrepancy for that observable, the discrepancy associated with the other 
parameters becomes significant.  More precise measurements of all of 
these parameters are required in order to clarify if there is an underlying
experimental issue that needs to be resolved, or if this is a sign of 
physics beyond the SM.

Prior to \superb starting to 
take data the \lhcb experiment should improve the precision on 
$\beta$ to about $0.5^\circ$, and on $\gamma$ by a factor of $2-3$ relative to the 
current state of the art, which may go some way to understanding the 
current tension in the constraints on the $bd$ unitarity triangle. 
The CERN based NA62 experiment will measure
the CKM parameter $\eta$ (the height of the triangle), using 
100 $K^+\to \pi^+ \nu\overline{\nu}$ events placing a new 
constraint on the height of the $bd$ unitarity triangle~\cite{NA62}.

\superb will be able to reduce uncertainties 
on $\alpha$, $\beta$,and $\gamma$ to the level of $1^\circ$, $0.1^\circ$, and $1^\circ$,
respectively to facilitate a precision CKM determination, and 
also improve the precision with which \Vub\ and \Vcb\ are measured.
The latter two observables are discussed below.
As pointed out in Ref.~\cite{Bevan:2011up}, it will be necessary to improve
constraints on $\Delta\Gamma_{B_d}$ in order to achieve these goals.
The set of precision CKM angle constraints can be used to test the 
SM description of quark mixing and CP violation to the level of 1\%.
SuperB will also produce a precision measurement of the branching fraction
of $\B \to \tau \nu$, and thus will also be able to probe the issue 
of current tensions observed in the SM.  
One should not forget that the process of making a precision test of the 
CKM matrix through measurement of these observables also improves the 
SM reference point that many other NP searches require.
Thus it is imperative that \superb performs
precision direct and indirect measurement of the $bd$ unitarity triangle
outlined here.

The most interesting \B decays that can manifest \CP violation at \superb 
are dominated either by tree or loop (penguin) transitions.  A number of these are known
to be theoretically clean, and calculations of SM uncertainties for a number 
of the other modes can be improved over the coming decade.  Time-dependent
\CP asymmetries are given by 
\begin{eqnarray}
{\cal A}(\deltat) &=& -C \cos \Delta M \deltat + S \sin\Delta M \deltat,
\end{eqnarray}
where $\Delta M$ is the $\Bz-\Bzb$ mixing frequency, $\deltat$ is the proper time
difference between the decay of two correlated \Bz mesons produced in \FourS collisions,
and both $S$ and $C$ are parameters related to \CP violating effects.  These are
discussed in detail in Ref.~\cite{Bevan:2011up}.
Collectively the measured differences in the antisymmetric \CP asymmetry, 
parameterised by $S$, in penguin decays and 
the benchmark $\Bz\to J\psi K_S$ channel are known as $\Delta S$ measurements.
In addition to the SM penguin amplitude new heavy particles could contribute 
additional amplitudes to these final states, and the interference between SM
and NP contributions could be detectable as an observable deviation from the tree $\sin 2\beta$
value.  More recently however the focus of these measurements has been extended
to compare the tree measurement of $\sin 2\beta$ against the inferred indirect
constraint on this parameter.  Thus this class of time-dependent \CP violation
mode serves as a set of sensitive interferometers for NP contributions from both tree and loop
amplitudes.  The modes under study, and corresponding theoretical and experimental
sensitivities achievable are listed in Table~\ref{tbl:deltas} (reproduced
from Ref.~\cite{physicswp}).  The channels are grouped into common physical final
states: $b\to c\overline{c}s$ charmonium decays, $b\to s$ penguin dominated
decays, and $b\to d$ penguin dominated decays.

\begin{table*}[!ht]
\caption{Current experimental precision of $S$~\cite{hfag}, and that expected at a \superb experiment with 75\invab of data.  
The $3\sigma$ and $5\sigma$ discovery 
limits at 75\invab are also listed.  The first entry in the table
corresponds to the tree level reference mode, and the next two sections of the table
 refer to $b\to s$
and $b\to d$ transitions, respectively. Theoretical estimates of $\Delta S$ are taken from Refs.~\cite{Beneke:2005pu,Cheng:2005bg,Cheng:2005ug}.
A long dash `$-$' denotes that there is no theoretical estimate of $\Delta S$ computed yet for a given mode,
thus the corresponding discovery limits are not evaluated.}\label{tbl:deltas}
\begin{center}
\small
{
\begin{tabular}{l|ccc|ccc|cc}\hline
Mode                & \multicolumn{3}{c}{Current Precision} & \multicolumn{3}{|c|}{Expected Precision (75\invab)} &
\multicolumn{2}{c}{Discovery}\\
  & &&& &&& \multicolumn{2}{c}{  Potential} \\ 
                    & Stat.        & Syst.       & $\Delta S^f$(Th.) & Stat.        & Syst.       & $\Delta S^f$(Th.)  & $3\sigma$ & $5\sigma$ \\ \hline
$J/\psi K^0_S$      & 0.018        & 0.009       & $0\pm0.01$        & 0.002        & 0.005       & $0\pm0.001$        & 0.02  & 0.03 \\ \hline
$\eta^\prime K^0_S$ & 0.08         & 0.02        & $0.015\pm0.015$   & 0.006        & 0.005       & $0.015\pm0.015$    & 0.05  & 0.08\\
$\phi K^0_S \pi^0$  & 0.28         & 0.01        & $-$               & 0.020        & 0.010       & $-$                & $-$   & $-$\\
$f_0 K^0_S$         & 0.18         & 0.04        & $0\pm0.02$        & 0.012        & 0.003       & $0\pm0.02$         & 0.07  & 0.12\\
$K^0_SK^0_SK^0_S$   & 0.19         & 0.03        & $0.02\pm0.01$     & 0.015        & 0.020       & $0.02\pm0.01$      & 0.08  & 0.14\\
$\phi K^0_S$        & 0.26         & 0.03        & $0.03\pm0.02$     & 0.020        & 0.005       & $0.03\pm0.02$      & 0.09  & 0.14\\
$\pi^0 K^0_S$       & 0.20         & 0.03        & $0.09\pm0.07$     & 0.015        & 0.015       & $0.09\pm0.07$      & 0.21  & 0.34\\
$\omega K^0_S$      & 0.28         & 0.02        & $0.1\pm0.1$       & 0.020        & 0.005       & $0.1\pm0.1$        & 0.31  & 0.51\\
$K^+K^-K^0_S$       & 0.08         & 0.03        & $0.05\pm0.05$     & 0.006        & 0.005       & $0.05\pm0.05$      & 0.15  & 0.26\\
$\pi^0\pi^0 K^0_S$  & 0.71         & 0.08        & $-$               & 0.038        & 0.045       & $-$                & $-$   & $-$\\ 
$\rho K^0_S$        & 0.28         & 0.07        & $-0.13\pm0.16$    & 0.020        & 0.017       & $-0.13\pm0.16$     & 0.41  & 0.69\\\hline
$J/\psi \pi^0$      & 0.21         & 0.04        & $-$               & 0.016        & 0.005       & $-$                & $-$   & $-$\\
$D^{*+}D^{*-}$      & 0.16         & 0.03        & $-$               & 0.012        & 0.017       & $-$                & $-$   & $-$\\
$D^{+}D^{-}$        & 0.36         & 0.05        & $-$               & 0.027        & 0.008       & $-$                & $-$   & $-$\\ \hline
\end{tabular}
}
\end{center}
\end{table*}

While it is also possible to measure direct \CP asymmetries in a large number of modes
at \superb, in general these are of limited use in terms of constraining theory.  
One exception is the \CP asymmetry in $b\to s \gamma$ decays discussed above.
The fundamental 
problem is that many of these measured observables are not theoretically clean, thus 
it is difficult to translate a direct \CP measurement into an unambiguous constraint 
on a SM parameter.
This is the $B$ physics analog of the issue associated with interpreting 
the measurement of direct \CP violation in kaon decays, $\epsilon^\prime/\epsilon$,
beyond establishing that such an effect exists, which in itself was an important 
goal.  It may be possible to combine many 
Charmless \B decay modes to test the SM using direct \CP asymmetries, however such tests
will probably never be as clean a set of observables as some of the time-dependent \CP asymmetries
discussed previously.  A prime example of this situation can be seen in terms of the 
difference between the direct \CP asymmetry measurements in $B\to K \pi$ decays.  
Some authors have advocated that the discrepancy is clear evidence for NP, however over
the last few years there have been a number of SM-based theoretical calculations 
that are able to explain this phenomenon.  For this reason it is unlikely that direct 
\CP violation measurements in Charmless hadronic \B decays will play a 
leading role in future experiments.  In Ref.~\cite{Gronau:2005kz} Gronau proposed a 
sum rule that could be used to correlate the measured asymmetries in $K\pi$ decays.

As mentioned above, it is also important to measure the sides of the $bd$ unitarity triangle,
using semi-leptonic decays.  The motivation for these measurements is two-fold: firstly
to resolve the experimental discrepancy between inclusive and exclusive measurements
left as a legacy of \babar and \belle, and secondly to try and resolve the current tension
between $\sin2\beta$, \vub, and the branching fraction of the decay $\B\to\tau\nu$.


The uncertainty on the indirect constraint of the location of the apex of the
$bd$ unitarity triangle is dominated by the experimental constraint on \Vub.  The
constraint obtained from $b\to u\ell \nu$ decays on \Vub\ has a precision of $\sim 11\%$,
whereas measurements of \Vcb, \Vcd, and \Vcs have uncertainties between $3$ and $5\%$.  
In the longer term, \superb is expected
to be able to improve the precision on \Vub\ to $2 (3)\%$ for an
inclusive (exclusive) measurement.  The precision on \Vcb obtained using 
$b\to c\ell\nu$ decays could be improved
from the current level of $3.5\%$ to $\sim 1\%$ for both inclusive and exclusive
measurements.  The increase in the precision of \Vub\ and \Vcb\ require 
some improvement in the precision of Lattice QCD input parameters.
The remaining quantities, \Vcd\ and \Vcs, can be measured using the charm
decays $D\to \pi \ell \nu$, and $D_s\to \ell \nu$, respectively.  It is likely
that the most precise measurements that can be made of these quantities at
\superb will use data accumulated at $\D\Db$ and above $\D_s$ thresholds.

The potential precisions outlined here require improvements in Lattice QCD,
that have been predicted up to 2015, and these are discussed in detail 
in Refs.~\cite{Bona:2007qt,physicswp}.
Until now all but one of these expected improvements in precision of 
Lattice quantities has proceeded at the anticipated rate, and in many case surpassed~\cite{physicswp}.  
Thus it is expected that future improvements in Lattice QCD will be made by the 
time that \superb starts taking data, and hence that the estimates discussed
in this section will be achievable.

\subsection{B physics at the \FiveS}
\label{bphysics:5S}

One of the motivations of studying the $B_s$ system at \superb is that the 
$\epem$ environment is extremely clean, so decays involving neutrinos or
many neutral particles, that would be inaccessible to an experiment at a
hadron collider, can be studied in detail.  It is not possible to study
$B_s$ mixing or time-dependent asymmetries in the $B_s$ system at existing or proposed \epem\ colliders 
because of the large mixing frequency, $\Delta m_s$.  Current detector technology 
would be unable to resolve oscillations in this decay, without having an extreme
boost for the center of mass system, relative to the laboratory frame of reference.
However it will be possible
to measure a number of interesting decays, including the semi-leptonic 
asymmetry $a_{SL}$ discussed in Section~\ref{bphysics:5S:asl}, and 
$B_s \to \gamma\gamma$ (see Section~\ref{bphysics:5S:gg}).  In addition, 
the increased knowledge of branching fractions 
obtained via measurements at $\epem$ facilities will help improve the precision of
absolute rates of $B_s$ decay modes, as \lhcb reports branching ratio measurements,
as opposed to branching fractions and will be limited by the absolute results given in the PDG
for the normalisation modes used.
More details on the \FiveS programme at \superb can be found in Ref.~\cite{physicswp}.

\subsubsection{Semi-leptonic asymmetry}
\label{bphysics:5S:asl}
$\,$

The semi-leptonic asymmetry measured in $B_s$ decays is of potential interest for NP searches. The asymmetry itself
if given by
\begin{eqnarray}
A_{\mathrm{SL}}^s = \frac{{\cal B}(B_s\to \overline{B}_s \to X^-\ell^+ \nu_\ell) - {\cal B}(\overline{B}_s\to B_s \to X^+\ell^- \nu_\ell)}{{\cal B}(B_s\to \overline{B}_s \to X^-\ell^+ \nu_\ell) + {\cal B}(\overline{B}_s\to B_s \to X^+\ell^- \nu_\ell)} = \frac{1 - |q/p|^4}{1 + |q/p|^4}.\nonumber
\end{eqnarray}
While 
this can be measured in hadronic environments, there is an intrinsic charge asymmetry that needs
to be understood, and controlled to high precision~\cite{lhcbasym}.  One way for hadronic experiments
to control this factor is to measure the difference in asymmetries between $B_d$ and $B_s$ decays 
$\Delta A^{d,s} = A_{\mathrm{SL}}^d-A_{\mathrm{SL}}^s$.  The corresponding measurement in an
\epem environment would enable a direct measurement of $A_{\mathrm{SL}}^s$ with smaller systematic
uncertainties, as well as having a different production environment that could be useful
in order to understand any deviations from SM expectations obtained.  
The anticipated precision for a measurement of $A_{\mathrm{SL}}^s$ with 1\invab of
data at \superb is 0.006.  It would also be possible to make an inclusive measurement of the 
asymmetry for both $B_s$ and $B_d$ decays, $A_{CH}$, with a precision of 0.004.  The current measurement
of the semi-leptonic asymmetry measured for a combination of $B_d^0$ and $B_s^0$ mesons 
from the D0 experiment is $3.9\sigma$ from SM expectations~\cite{Abazov:2011yk}, where the
asymmetry $A_{\mathrm{SL}}^b = (-0.78\pm 0.17\pm 0.09)\%$, and one expects $(-0.028\pm 0.005)\%$ in the SM.

\subsubsection{$B_s \to \gamma\gamma$}
\label{bphysics:5S:gg}
$\,$

The SM branching fraction for $B_s \to \gamma\gamma$ is expected to be 
$0.5-1.5\times 10^{-6}$ (see for example see Ref.~\cite{Huo:2003cj}, and references therein),
and the decay proceeds via a $b\to s\gamma\gamma$ FCNC loop transition.  Hence the NP that can affect 
possible $b\to s$ penguin measurements of time-dependent \CP asymmetries or other kinematic quantities 
may also be at play
in this decay.  This channel benefits from the lack of hadronic particles in the final 
state, when compared to the $\Delta S$ measurements discussed in Section~\ref{bphysics:4S:angles},
and is also related to similar $B_d$ channels.

This is an experimentally challenging final state to isolate and extract, and it is only possible
to isolate this channel at an \epem collider based experiment.  The two-photon invariant
mass distribution will have a significant background from high energy combinatoric photons,
as has been seen in previous searches for this decay~\cite{Wicht:2007ni}.  The current upper 
limit for this channel is $<8.7\times 10^{-6}$ at 90\% C.L. obtained by \belle from a 
data sample of 23.4\invfb recorded at the \FiveS. However care should be taken when extrapolating
this number to higher luminosities as this limit is obtained from a downward fluctuation,
resulting from a slightly negative event yield obtained from the fit to data.
The challenge here for \superb
is to isolate as clean a signal as possible, using only information from the EMC to provide 
a positive identification for $B_s \to \gamma\gamma$, and the other sub-detectors as an
elaborate veto system.  The other $B$ meson in the final state can be used to provide sufficient
kinematic information to help reduce the background level.  The SM rate should be attainable
at \superb where it is expected that sensitivities of the order of $3\times 10^{-7}$ can be reached.
Thus if the signal were manifest at the upper end of expectations, \superb should be able to 
observe it, and if the true branching fraction were at the lower end, then it would be 
challenging to establish evidence for the existence of this decay. 

The $\gamma\gamma$ branching fraction can be affected by NP scenarios in a similar way to
$B_{d,u} \to X\gamma$, $\gamma\gamma$, and $X\gamma\gamma$, where significant enhancements
above the SM rate are possible~\cite{PhysRevD.70.035008,Bertolini:1998hp}.
In addition to the scenarios where the $B_{d,u}$ decays are correlated with 
$B_s \to \gamma\gamma$, there are also specific models where there is no correlation and 
this decay can be significantly enhanced by NP, whereas the $B_{d,u}$ counterparts remain 
un-affected~\cite{Huo:2003cj,Aranda:2010qc}. 
This makes $B_s \to \gamma\gamma$ an important decay to measure
in order to provide an independent cross check of any deviation observed in a $B_{u,d}$ mode.  

%% file: dphysics.tex
\section{\boldmath{D physics}}
\label{dphysics}

Charm analyses at \superb are broadly split into two categories, those 
using data collected at the \FourS, and those collected at or near charm 
threshold, the $\psi(3770)$.  The \FourS data results from $D$ mesons 
being produced from \epem continuum events, where the cross-section for 
$c\bar c$ is comparable to that for $b\bar b$.  
In general one reconstructs tagged \D mesons from a $D^* \to D \pi_s$ transition
where $\pi_s$ denotes a slow (low-momentum) pion.
The advantage
of studying charm in \superb at this energy is the vast data sample that can be
collected in a clean environment where one expects to accumulate
$90\times 10^{9}$ $\D^0$ meson pairs in 75\invab of data.  The drawback is that the data,
while clean, are not background free, and for example one must restrict the momentum range
of $D$ mesons to exclude events originating from $B$ decay.  In some 
measurements systematic uncertainties from background may be a critical
issue, and for these having access to data collected at the $\psi(3770)$ resonance 
may provide a distinct advantage.  In addition to having smaller background,
neutral $D$ meson pairs produced at charm threshold are quantum correlated where
one always has a $\Dz$ and a $\Dzb$ until one of the mesons decays.
In essence one can repeat the $B$ factory experiment at the \FourS, with a $D$ 
factory experiment at charm threshold.  In order to exploit the full potential of 
the quantum correlated neutral meson pair one needs to have a boosted centre of mass
system, and this may be achievable with $\beta\gamma$ up to $0.91$ at \superb. 
The baseline boost for \superb is currently somewhat smaller than this value.
The drawback of running at charm threshold, with respect to the \FourS, 
will be that the accumulated luminosity at the $\psi(3770)$ is expected to only be
of the order of 500\invfb.  This will result in only $\sim 1.8\times 10^{9}$
$D$ meson pairs being produced, however these data 
are extremely clean, and kinematics of the initial state $\epem$ pair and the `other' $D$ meson
in the event can be used to essentially select samples of almost pure $D$ mesons.  A number of observables measured
using the data collected in a few months at the $\psi(3770)$ will be competitive 
with results from the \FourS sample accumulated over the lifetime of \superb, and some will
help control systematic uncertainties in measurements made using \FourS data at \belletwo
and \superb, and also help reduce uncertainties for the corresponding measurements
at \lhcb. Many observables can be accessed using both samples of data, and 
are discussed according to topic in the following.

\subsection{Charm Mixing}
\label{dphysics:mixing}

The last largely uncharted area of neutral meson mixing that remains to 
be explored is that of the charm sector.  The \B Factories 
found evidence for charm mixing in 2007 using studies of 
$D\to K^+\pi^-$ decays~\cite{Aubert:2007wf}, and subsequently using 
$D\to h^+h^-$ decays~\cite{Staric:2007dt} (where $h=\pi, K$), and have 
started the search for time-integrated \CP violation in charm transitions. 
While both neutral \B and \K mesons have been studied in detail, one
should recall that these involve flavour changing transitions of down 
type quarks.  The study of mixing and \CP violation in the charm 
sector corresponds to the study of an up-type quark, where any large 
manifestation of \CP violation would constitute a sign of NP.
It should also be noted that as the amount of data accumulated increase,
additional observables will become accessible to experimentalists.  An
example of this very situation can be seen in terms of time-dependent
\CP asymmetry measurements, discussed in Section~\ref{dphysics:cpvtd}.

\begin{figure}[!ht]
\begin{center}
  \includegraphics[width=10cm]%
        {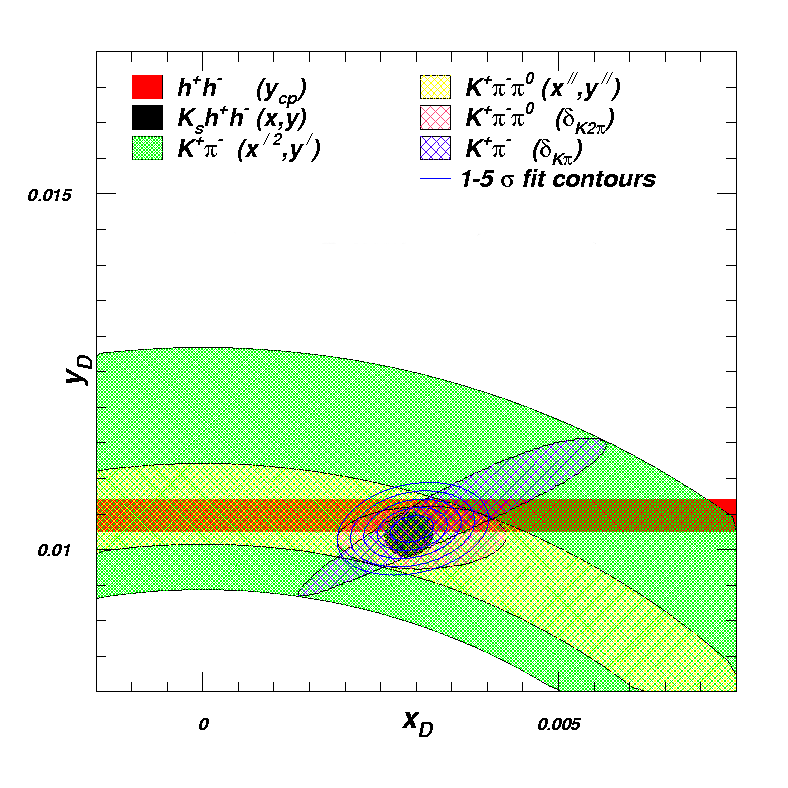}
  \caption{The constraints obtained on the charm mixing parameters $\xd$ and $\yd$ using \FourS and $\psi(3770)$ data
    from \superb (Ref.~\cite{physicswp}).}
  \label{fig:dphysics:mixing}
\end{center}
\end{figure}

Neutral $D$ mesons mixing can be described by a Hamiltonian consisting 
of a $2\times 2$ matrix of elements given by $\boldmath{M} + i\boldmath{\Gamma}/2$,
where $\boldmath{M}$ and $\boldmath{\Gamma}$ are themselves $2\times 2$ matrices~\cite{bigiandsanda}.
The weak eigenstates of neutral $D$ mesons can be 
expressed as admixtures of the strong states
\begin{eqnarray}
|D_{1,2}\rangle = p|D^0\rangle \pm q|\overline{D}^0\rangle,
\label{eq:tdep:admixture}
\end{eqnarray}
where $q^2+p^2=1$.  The characteristic mixing frequency is given by $\Delta M$,
which is given by the mass difference of the weak eigenstates.  The other relevant observable
is $\Delta \Gamma$, given by the width difference of those eigenstates.  Experimentally
one measures mixing via the parameters $\xd$ and $\yd$ (or related parameters) where
\begin{eqnarray}
\xd = \frac{\Delta M}{\Gamma},\,\,\,\,\mathrm{and} \,\,\,\, \yd = \frac{\Delta \Gamma}{2\Gamma}.
\end{eqnarray}
Given the small values of $\xd$ and $\yd$ an approximation is used for the time-dependence of the
evolving neutral meson state including only quadratic and linear terms of these parameters.
A further complication enters the measurement as in general all final states $f$ have a
relative strong phase $\delta_f$ (invariant under \CP) that needs to be determined, neglecting the weak
phases that are expected to be small in the SM.  Hence in general one 
measures parameters in a rotated basis that are related to $\xd$ and $\yd$ given by
\begin{eqnarray}
x^\prime_f &=& \xd \cos \delta_f + \yd \sin\delta_f,\nonumber\\
y^\prime_f &=& \yd \cos \delta_f - \xd \sin\delta_f.\label{eq:dphysics:xandy}
\end{eqnarray}
A number of final states have been studied in order to determine the $D$ meson 
mixing parameters, these include wrong sign $D\to K\pi$~\cite{Aubert:2007wf,Aaltonen:2007uc}, 
$hh$~\cite{Staric:2007dt,Aubert:2009ck}, $K\pi\pi^0$~\cite{Aubert:2008zh} 
and $\KS hh$~\cite{Abe:2007rd,delAmoSanchez:2010xz} decays.  
Experimentally $\xd$ and $\yd$ are found to be small, where $x\sim 0.005$ and $y\sim 0.01$~\cite{hfag},
thus both $\Delta M$ and $\Delta \Gamma$ are small for neutral $D$ mesons.
The dominant contribution to these measurements currently
comes from the time-dependent Dalitz Plot (DP) analysis of $\Dz \to \KS h^+h^-$ final states as one is able
to determine the value of $\delta_f$ as a function of position in the DP, and hence extract $\xd$ and $\yd$ directly
for this mode. The other channels measure $\xd$ and $\yd$ up to a rotation corresponding to the strong phase 
measured in the final state according to Eq.~(\ref{eq:dphysics:xandy}).
There is an intrinsic limit to the precision of any measurement using the $\KS h^+h^-$ 
final state that comes from the DP model used
for the decay.  It will be possible to control this contribution to a charm mixing
analysis by performing a detailed study of this decay using data collected at
charm threshold (see Section~\ref{sec:dphysics:threshold}).  Table~\ref{tbl:dphysics:mixing}
summarises the expected precisions obtainable on mixing parameters using existing methods and data from \superb.
The impact of charm threshold running on the determination of these parameters is clearly 
evident.  On inclusion of the improved DP information one will be able to halve the total
uncertainty on $\xd$ and reduce the uncertainty on $\yd$ by 30\% and Figure~\ref{fig:dphysics:mixing} shows
the different constraints obtainable using all data from \superb.  More importantly
these results will change from being systematically to statistically limited,
allowing for further improvements.  The \superb results
from threshold running will impact upon the ultimate precision of mixing parameters determined by 
the \belletwo and \lhcb experiments.
It will also be possible to place model dependent constraints on $|q/p|$ and the phase of charm 
mixing with precisions better than $1-2\%$ and $1.4^\circ$, respectively.

\begin{table}[!h]
\caption{Expected precision on charm mixing parameters using \FourS data from \superb
with existing methods from the \B factories.  The estimates given in the first two rows
include only data from the \FourS, while the results in the last two rows
combine the \FourS expectations with an improved $\KS\pi\pi$ DP model resulting from 
a charm threshold run.}
\label{tbl:dphysics:mixing}
\begin{center}\begin{tabular}{lcccc}\hline
Parameter     & $x\times 10^{3}$  & $y\times 10^{3}$ & $\delta_{K\pi}$ $(^\circ)$& $\delta_{K\pi\pi}$ $(^\circ)$\\ \hline  
$\sigma$ \stat           & 0.18 & 0.11 & 1.3 & 2.7\\
$\sigma$ \stat+\syst     & 0.42 & 0.17 & 2.2 & $\,^{+3.3}_{-3.4}$\\ \hline
$\sigma$ \stat           & 0.17 & 0.10 & 0.9 & 1.1\\
$\sigma$ \stat+\syst     & 0.20 & 0.12 & 1.0 & 1.1\\
 \hline
\end{tabular}
\end{center}
\end{table}

An alternate method of studying charm mixing has been proposed using time-dependent \CP asymmetry 
measurements which is discussed in Section~\ref{dphysics:cpvtd}. One will be able
to measure the phase of mixing with a statistical precision of ${\cal O}(1.3^\circ)$ using 
$D^0\to K^+K^-$, and one may be able to achieve sub-degree level measurements 
using more copious decays such as $K_S\pi^0$~\cite{Bevan:2011up}.

\subsection{Time-dependent \CP violation}
\label{dphysics:cpvtd}

It is possible to perform time-dependent \CP asymmetry measurements in 
$D$ decays, both at threshold and using data collected at the \FourS.
The motivation and formalism for doing so has been discussed recently
in Ref.~\cite{Bevan:2011up} in the context of testing the $cu$ unitarity 
triangle.  An important issue to raise is that in order to understand the 
phenomenology of \CP violation one has to choose a convention for the 
four parameters of the CKM matrix, and the order to which one expands
the matrix elements in terms of this basis.
Until recently the Wolfenstein parameterisation of the CKM matrix~\cite{Wolfenstein:1983yz}
has been used by default.  Here the four expansion parameters of the matrix are
$\lambda = \sin\theta_c \sim 0.22$, $A$, $\rho$ and $\eta$, and the description of the matrix has
been given up to ${\cal O}(\lambda^3)$ in Eq.~(\ref{bevan:eq:ckm}).  An alternative parameterisation 
using the same four parameters has been proposed by Buras et al. in Ref.~\cite{Buras:1994ec}
which has the advantage that the unitarity triangle is unitary to any order 
of the expansion in terms of $\lambda$.
At ${\cal O}(\lambda^3)$, both conventions are equivalent, however in the 
charm sector one needs to consider additional terms, to at least 
${\cal O}(\lambda^5)$, to understand the CKM structure and \CP violation
potential of charm decays.  

The charm `$cu$' triangle given by
\begin{eqnarray}
\vud^*\vcd + \vus^*\vcs + \vub^*\vcb = 0, \label{eq:charmtriangle}  
\end{eqnarray}
has been 
known for some time~\cite{Bigi:1999hr}, however thus far there have been no direct tests
of unitarity for this triangle.
Ref.~\cite{Bevan:2011up} outlines the procedure to measure the 
mixing phase using time-dependent \CP asymmetries of \CP eigenstates such as 
$D\to K^+K^-$ and one of the angles of the $cu$ triangle, $\beta_c$, using
a combination of $D\to K^+K^-$ and $D\to \pi\pi$ final states.  The phase
difference measured between these two modes is related to the the observable
$-2\beta_{c,\mathrm{eff}}$ measured in the time-dependence of $\Dz\to \pi^+\pi^-$ decays.
The subscript `$\mathrm{eff}$' indicates that while this parameter is related to
the angle $\beta_c$ of the $cu$ triangle, there are theoretical uncertainties that 
may cause $\beta_c$ to differ from $\beta_{c,\mathrm{eff}}$.  Ref.~\cite{Bevan:2011up} discusses
this issue, highlighting in particular that one can perform an Isospin analysis of 
$D\to \pi\pi$ final states in order to disentangle
the effect of penguin amplitudes that contribute to $\Dz\to \pi^+\pi^-$ with 
a different weak phase than the leading tree contribution.  As this Isospin
analysis requires input from $\Dz\to \pi^0\pi^0$ and $\D^+\to \pi^+\pi^0$,
it will not be possible to perform a self-consistent measurement of 
$\beta_{c,\mathrm{eff}}$, correcting for penguins in a hadronic environment.
While hadron experiments may provide input to help solve this problem 
one will require data from a \sff to complete the picture.  It has been 
estimated that it will be possible to constrain $\beta_{c,\mathrm{eff}}$ to a 
precision of $1^\circ$ at \superb before taking into account penguin
contributions~\cite{Bevan:2011up} on combining data from threshold and the \FourS.  
As $\beta_c$ is estimated to be $\sim 0.035^\circ$, \superb will be able to constrain
large NP effects in time-dependent \CP asymmetry measurements in charm
decays, but will lack the precision to perform a direct test 
of the SM.  Nonetheless, it will be important to verify that $\beta_{c,\mathrm{eff}}$
is consistent with zero.  One should also note that time-dependent measurements
will also be able to provide model dependent constraints on $|q/p|$,
and so one can search for both direct and indirect \CP violation, in addition to 
measuring the phase of neutral $D$ meson mixing.  

Some of the systematic uncertainties in time-dependent \CP asymmetry measurements
made using data collected at the $\psi(3770)$ resonance will be different
from those at the \FourS.  This provides a potential advantage to \superb
as one can perform an independent cross check of any phase measurement
for consistency.  While the precision of $\beta_{c,eff}$ using 500\invfb
of data from the $\psi(3770)$ is expected to be slightly worse than 
that from 75\invab at the \FourS, the average of these two results may
approach the level of $1^\circ$.  Studies are ongoing in this area.

In addition to direct tests of the CKM matrix via measurements of an
angle of the $cu$ triangle, one can indirectly test this using constraints
on the magnitudes of \vub, \vcb, \vus, \vcb, and \vcs.  In principle these
quantities can be measured using a combination of data collected at the 
\FourS (\vub\ and \vcb), in the vicinity of the $\psi(3770)$ (\vcb\ and \vcs), and any energy
using $\tau$ decays (\vus).  The other CKM matrix element that is required
as an input for an indirect side constraint is \vud, however this is already
precisely known~\cite{Nakamura:2010zzi}.
 
As with \B decays it is possible to search for direct \CP violation in charm decays where
the interference between two or more such amplitudes can manifest a \CP asymmetry.  It is expected
that the direct \CP asymmetry in $D\to hh$ decays could be as large as $10^{-4}$, which would be
manifest in a time-dependent \CP asymmetry measurement.  As is the case with $K$ and $B$ decays,
different strong phases are required in order to manifest a non-trivial effect, and as these
are difficult to theoretically calculate, any measurement of a small level of direct \CP violation in charm would
most probably be of limited use in testing the SM beyond establishing the existence of such an effect.

\subsection{Rare decays}
\label{dphysics:rare}

Rare \D decays are sensitive to NP scenarios, and can provide an important test of the SM.  A number
of possible NP probes are being studied however it is clear that there are several important measurements
to be made at \superb, including searches for $D\to \gamma\gamma$, $\ell^+\ell^-$, 
and $\nu\overline{\nu}(+\gamma)$.

\begin{description}
\item {\boldmath{$D\to \ell^+\ell^-$}}: 
  The SM expectation of the branching fraction of $D\to \mu^+\mu^-$ is dominated by 
  a long-distance contribution which is related to the $D^0\to \gamma\gamma$ rate via
\begin{eqnarray}
{\cal B}(D\to \mu^+\mu^-)_{LD} = 3.0\times 10^{-5} \cdot {\cal B}(D^0\to \gamma\gamma).
\end{eqnarray}
  The expectation is that this decay proceeds at the level of $3\times 10^{-13}$ which can be inferred
  from the expected rate of $D^0\to \gamma\gamma$ discussed below.
  Significant enhancements to the branching fraction can be obtained in models of NP.  
  The current experimental 
  bound is $<1.4\times 10^{-7}$~\cite{Petric:2010yt} which is about an order of magnitude larger
  than possible enhancements from R parity violating SUSY.  \superb should be able to improve 
  upon these limits and reach a sensitivity an order of magnitude better than the current 
  constraints, and in doing so may start to constrain NP parameter space, however it is likely
  that \lhcb will be able to place a more stringent constraint on this mode.  Having measured the
  branching fraction from data, one is limited in terms of interpretation of this result in the
  context of NP by the lack of knowledge on the long-distance rate.  The related channel discussed
  below can help elucidate this situation, and is an example of the natural synergy between
  hadron and \epem environments.

  The di-electron mode is also of interest, however it is more difficult to trigger on electrons
  in a hadronic environment.  The \sffs will produce competitive limits on ${\cal B}(D\to e^+e^-)$.
  Currently the most stringent limit on this decay is $7.9\times 10^{-8}$ from \belle~\cite{Petric:2010yt}.

\item {\boldmath{$D^0\to \gamma\gamma$}}: The two-photon channel is expected to have a 
  branching fraction of $(1.0\pm 0.5)\times 10^{-8}$ in the SM~\cite{Burdman:2003rs}.  The current 
  experimental limit~\cite{Coan:2002te} on this decay comes from CLEO
   and is $<2.9\times 10^{-5}$.   An improved limit on this channel can be used to constrain 
   the long-distance contribution to the di-lepton final states.  It is expected that \superb will be 
   able to achieve a sensitivity of a few $\times 10^{-7}$ in this channel, and while an order of 
   magnitude larger than the SM expectation, this result would be able to constrain possible 
   enhancements in the $D\to \mu^+\mu^-$ channel at a useful level.
   The corresponding limit on ${\cal B}(D\to \mu^+\mu^-)_{LD}$ obtained from \superb
   could reach the expected SM level for this decay, i.e. $3\times 10^{-13}$.  Thus by combining
   results from hadron experiments on $D\to \mu^+\mu^-$ with a limit on $D^0\to \gamma\gamma$
   from an \epem experiment, one will be able to search for NP.

\item {\boldmath $D\to \nu\overline\nu (\gamma)$}: Decays of heavy mesons into invisible final states, or $\gamma +$invisible
     states can be used to probe for signs of light Dark Matter~\cite{Badin:2010uh}.  The SM decay into $\nu\overline\nu$
     is helicity suppressed, hence any signal found would provide a clear indication of NP. 
     Such a measurement at \superb would complement the corresponding studies performed in 
     $B$ and $\Upsilon$ decays.  In order to perform such an analysis one would have to use data collected at charm threshold,
     so that the kinematics of the final state $\Dz\to \nu\overline\nu$ can be constrained by measurement of the recoil
     $D$ meson and knowledge of the initial $\epem$ kinematics in the decay chain $\epem \to \psi(3770) \to \Dz\Dzb$.  Unlike
     $B$ decays where the sum of branching fractions for fully reconstructed final states is a few percent, here a $D$
     recoil analysis would utilise over half of the available final state $D$ decays.  With a total of $1.8\times 10^{9}$
     $D$ mesons produced at threshold it is feasible to assume that \superb will be able to perform a detailed search for 
     both of these decays. While the implied single event sensitivity would be $({\cal O})(few \times 10^{-9})$, one should 
     expect that there might be a significant level of residual background resulting from the lack of hermiticity of the 
     detector, but any background would be less than that found in the corresponding searches for $B$ to invisible final 
     states.  The sensitivity achievable is under study, and it is likely that $\Dz\to \nu\overline\nu$ will suffer
     from an irreducible background from $D\to K\pi$ decays, where the final state particles go down the beam pipe.
     There is a similar interest in searching for $B_{d,s}\to invisible (+\gamma)$
     decays where the SM and light Dark Matter expectations are also discussed in ~\cite{Badin:2010uh}.
\end{description}

\subsubsection{\CPT with charm}
\label{dphysics:cpt}
$\,$

\CPT can be tested using decays of pairs of neutral $K$, $D$ and $B$ meson created 
in a quantum correlated state via decays at centre of mass energies corresponding
to the $\phi$, $\psi(3770)$, and \FourS respectively.  It is possible to 
perform a precise test of \CPT using $\Dz\Dzb$ pairs from the data sample collected 
at $\psi(3770)$, where both $D$ mesons decay into a semi-leptonic final state 
(e.g. see ~\cite{Kostelecky:2002pf}).
The interpretation of any potential \CPT violating effect is model dependent, so it is 
important to test this fundamental symmetry for all neutral meson systems.  Existing 
measurements are compatible with \CPT being an exact symmetry, however it is interesting
to note that the \babar experiment reported a $2.8\sigma$ deviation from \CPT conservation
when studying a large sample of di-lepton events as a function of sidereal time~\cite{Aubert:2007bp}.
As \superb expects to accumulate 50 times more statistics than BES III at threshold, it is expected 
that any constraint on \CPT produced would be a significant improvement over previous 
results, and complement the corresponding measurements made at the \FourS, as well as results
expected to be made by the KLOE-2 experiment over the next few years~\cite{AmelinoCamelia:2010me}.

\subsection{Other Measurements at Charm Threshold}
\label{sec:dphysics:threshold}
$\,$

\superb will have a dedicated run at the $\psi(3770)$ (charm threshold) in 
order to benefit from the extremely clean environment obtained by the use of D-recoil 
methods that partially or fully reconstruct the other D meson in the event,  and the quantum correlated $\Dz\Dzb$ system.
The power of identifying an almost pure sample of charm mesons at threshold to analyse can 
be seen by comparing the precision of many CLEO-c results to those from the
\B Factories.  In a number of cases, especially form factors, CLEO-c has been
able to out perform the \B Factories, and in some cases the measurements are unique to 
CLEO-c.

A total of 500\invfb of data will be collected at the $\psi(3770)$, and 
runs at adjacent resonances may also be performed to include samples of $D^+_{s}$ mesons.
In comparison CLEO-c accumulated $0.8\invfb$ of data at the $\psi(3770)$, and 
$0.6\invfb$ with a centre of mass energy of $4.17 GeV$.  BES III is expected to 
accumulate 10\invfb of data at threshold during the coming few years.  Hence
\superb is expected to accumulate 50 times the data of BES III and over 500 times the data
of CLEO-c.  This opens up the potential to cleanly search for, and measure a number of rare
decays that would otherwise be inaccessible.

A number of important measurements rely on a detailed understanding of charm decays.  One such 
example is the $D\to \K_S\pi\pi$ final state~\cite{Bondar:2008hh}, which feeds into both the measurement of $\gamma$
via the GGSZ method and traditional charm mixing analyses as discussed in 
Section~\ref{dphysics:mixing}.  There is an intrinsic model uncertainty associated with 
the use of this decay, which arises from the amplitudes considered in the DP model
and thus the values of strong phases extracted as a function of the DP.  CLEO-c have shown that 
one can use quantum correlated $D$ mesons produced at the $\psi(3770)$ in order to 
perform a measurement of the strong phase in $\K_S\pi\pi$ decays as a function of
position in the DP~\cite{Briere:2009aa}.  At the time this result came out, it was used
to halve the model uncertainty on $\gamma$ for the \B factory results.  In order for the 
GGSZ method to remain a viable approach for the measurement of $\gamma$ in the \sff era,
one will need an improved measurement of the Dalitz model for $\K_S\pi\pi$ decays.  As
mentioned in Section~\ref{dphysics:mixing}, the inclusion of this result will significantly impact
the measurements of charm mixing parameters at \superb, \belletwo and \lhcb, and
make future studies of these parameters meaningful. 
A number of other potential uses of charm decays to quantum correlated 
final states are under investigation within the \superb Collaboration.

%% file: precisionew.tex
\section{\boldmath{Precision electroweak decays}}
\label{precisionew}

As discussed in Section~\ref{sec:facility:accelerator}, one of the unique features of \superb is that the longitudinal 
polarisation of the electron beam. This enables one to study 
left-right asymmetries in \epem interactions that can be used to perform precision measurements
of $\sin^2\theta_W$.  There are two reasons why it is interesting to measure this parameter
at \superb; firstly this would provide a measurement of $\sin^2\theta_W$ at an energy of 10.58\gev, 
a region where the coupling is changing and there is no measurement so far.  Secondly the 
$\epem\to b\overline{b}$ measurement made at this energy would be devoid of the $b$ fragmentation
uncertainties that theoretically limit interpretation of results from SLC/LEP measurements.  The 
left-right asymmetry $A_{\mathrm{LR}}$ is constructed from measurements of the cross-section of
events with left and right helicities
\begin{eqnarray}
A_{\mathrm{LR}} = \frac{\sigma_{\mathrm{L}} - \sigma_{\mathrm{R}}}{\sigma_{\mathrm{L}} + \sigma_{\mathrm{R}}}
  = \frac{2a_v a_e}{a_v^2+a_e^2} = \frac{2[1-4\sin^2\theta_W^{eff}]}{1+[1-4\sin^2\theta_W^{eff}]},
\end{eqnarray}
where $a_v$ ($a_e$) is a vector (electron neutral current axial) coupling related to the decay.  The anticipated 
precision for $\sin^2\theta_W$ \superb is $\sim 2\times 10^{-4}$, and it should be possible to 
measure the $b-$quark vector coupling with a comparable precision to the SLC/LEP measurement~\cite{physicswp}.

There are other experiments either performing or proposing to make measurements of $\sin^2\theta_W$ in the coming years.
These include the JLab based QWeak experiment ($\sqrt{s}= 0.173 \gev$), and the LHCb upgrade ($Z$ pole).  
\belletwo does not have a polarised beam, and so it
won't be possible to make a measurement of $\sin^2\theta_W$, however that experiment will be able to 
improve our knowledge of the axial coupling related to this fundamental parameter by
measuring the forward-backward asymmetry.  This will be a useful input to the \sff precision electroweak
physics programme.

Improved measurements of $\sin^2\theta_W$ can be used to improve our understanding of precision electroweak predictions
based on the SM, or alternatively as constraints on scenarios of physics beyond the SM.  In particular precision measurements
are sensitive to models of NP with $Z^\prime$ bosons.

%% file: directsearches.tex
\section{\boldmath{Direct searches and exotica}}
\label{directsearches}

Most of the NP searches at \superb are indirect, where it is not possible to manifest 
the new particles in the laboratory.  However there are models where it would be 
possible to directly produce NP particles in low energy \epem interactions, and 
infer something about the type of new physics leading to their existence.  
This area is briefly discussed here, and more details can be found in ~\cite{physicswp}.
\begin{description}
 \item {\bf Dark Matter}
The expectation that Dark Matter exists is well known to be motivated by models
of the rotation of spiral galaxies, where significant amounts of undetected matter 
must exist in order to explain the visible part of the galaxies.  As a result there are
a number of experiments dedicated to searches for signs of the halo dark matter postulated
to exist in the Milky Way, some of which should be local to the Earth.  
While \superb is unable to contribute to searches for halo dark matter, it is possible
that small amounts of Dark Matter could be created in low energy \epem collisions.  
These would be manifest through the enhancement of rates for decays to final states with 
invisible particles such as those discussed in Section~\ref{dphysics:rare} and analogues in $B_d$ and $B_s$ decays.
 \item {\bf Dark Forces}
A relatively recent theoretical development is the scenario that there could be a scalar
field related to the so-called `Dark Sector'.  One of the predictions of these scenarios is
a \gev scale particle that decays into {\em dark photons}, which can subsequently affect
the kinematic distributions and rates of rare processes.  Experimental signatures that can
be used to search for evidence of the dark sector include meson decays into multi-lepton 
final states.  A recent review of these results can be found in Ref.~\cite{kolomensky}.
 \item {\bf Light Scalar Higgs}
The SM Higgs is known to have a mass above 114\gevcc, and both the Tevatron
and LHC are actively searching for the existence of such a particle.  However if or when the Higgs is found,
this particle itself introduces problems into the theory via self-coupling, and would motivate
some NP scenario that could involve the introduction of supersymmetric particles, or multiplets of 
Higgs particles.  In many scenarios of new physics with multiple Higgs', where one of these 
may be a light neutral particle that has not yet been ruled out by data from LEP and the \B Factories.  
In this scenario light means $<10\gevcc$.  This light scalar Higgs is denoted by $A^0$ 
is expected to decay predominantly into charged lepton pairs, where the most probably final 
state would be $\tau^+\tau^-$.  There have been a number of recent searches for such particles
by the \B factories~\cite{Lees:2011wb,Aubert:2009cka}.
\end{description}
Many direct searches for NP have been made by the current \B factories, and it will 
possible for \superb to make significant improvements on the limits obtained, where
for example one would typically assume an order of magnitude improvement on searches
for dark matter candidates.  The improvement on $A^0 \to \ell^+\ell^-$ transitions in
decays of $\Upsilon$ mesons will depend on the integrated luminosity obtained
for the various \NS resonances. 

\subsection{Lepton Universality}
$\,$

Using the same experimental signatures of light mesons $M^0$ decaying into di-lepton final states that are required
for light Higgs searches to test Lepton universality (LU).  In the SM the coupling strength associated
with lepton vertices is common and the branching fractions of some $M^0$ into a di-lepton final state
are equal up to factors related to the masses of leptons in the final state.  The set of measurements 
comparing ratios of branching fractions, corrected for the lepton mass-difference therefore provides
a measure of the lepton coupling, and any deviation from a common value could indicate a violation of 
LU.  As the lighter leptons can undergo bremsstrahlung, it is necessary to ensure that radiative effects
are properly accounted for when performing such measurements.  The results of recent tests of LU in 
$\OneS$ decays provides a test a the percent level~\cite{delAmoSanchez:2010bt}.  Tests at the sub-per mille
level using $\tau$ decays have also been reported~\cite{Aubert:2009qj}.

%% file: interplay.tex
\section{Interplay between measurements}
\label{interplay}

A priori we don't know the structure of physics beyond the SM, and we
only have model-based lower limits on the possible energy scale of
new particles based on naturalness arguments.  The 
basis of naturalness is to assume that couplings in a model of nature
are not fine tuned to small values, but may take arbitrary values
as large as ${\cal O}(1)$.  Such arguments are used set a scale of 
electroweak symmetry breaking at 1TeV.  On considering the historical
development of the SM as described in Section~\ref{intro:history},
it is also possible to use flavour changing processes to probe 
higher energies via the contribution of virtual effects to the
total amplitude of a rare process.  In the case of \Bz-\Bzb mixing 
a system with an energy of 5.28 GeV was used to detect the presence
of the top quark, which is now known to have a mass of $172\pm 0.9\pm 1.3$ 
$GeV/c^2$~\cite{Nakamura:2010zzi}.  One can perform a similar exercise
to constrain possible a NP energy scale $\Lambda_{NP}$, again using 
rare decays and flavour changing processes.  Such constraints are
model dependent, and depending on the model, the scale of new
physics can be placed between 10 and 100TeV.  An energy scale of 1TeV
or below can only be obtained by setting flavour parameters in 
the NP sector to zero.  Such models are known as Minimal Flavour
Violation (MFV) models.   The sources of CPV in a MFV model are 
the same Yukawa couplings from the Higgs sector that result in the 
CKM matrix. Given that there is a rich texture in
nature related to the flavour changing processes, for both 
quarks and neutrinos, it may seem improbable that any complex NP sector,
such as SUSY would be completely flavour blind, with all new \CP violating
phases arbitrarily set to zero.  Thus there
is a tension between well motivated naturalness arguments giving
an energy scale of 1TeV for the electroweak symmetry breaking,
and our expectation that new physics might have a rich texture related
to flavour changing processes in analogy with what we have observed
so far in the SM.  Experimental input is required in order to move
forward on both the high-energy and flavour fronts.  At the time
of writing this report, there has been no significant signature for 
NP encountered at the LHC in order to guide this exploration.  The
lack of a discovery has already started to have ramifications for 
flavour blind scenarios of NP.

In analogy with the pioneering work of Cabibbo in developing 
quark-mixing phenomenology~\cite{Cabibbo:1963yz}, if we are to
determine the structure of any underlying new physics scenario
that may provide a realistic description of nature at high energy,
we must combine constraints from a number of different measurements.
Thus in order to optimise our progress in this endeavor, we 
need to make as many independent measurements of theoretically
clean observables that might be affected by NP as possible.
Section~\ref{interplay:NPlagrangian} discusses some of the ways
currently envisaged to elucidate the structure of the NP Lagrangian
from rare processes based on both observed deviations from 
SM expectations and results consistent with the SM.  Section~\ref{interplay:NPlagrangian:massinsertions}
discusses the mass-insertion hypothesis and how one can relate
flavour observables to $\Lambda_{NP}$. The
three generation mixing matrix can be used as a reference point
to search new NP as discussed in Section~\ref{interplay:precisionCKM}.

\subsection{Reconstructing the new physics Lagrangian}
\label{interplay:NPlagrangian}

$\,$

The main purpose of \superb is to try and elucidate the structure of 
new physics at a level that goes beyond anything currently possible.
Not only can the existence of an unknown heavy particle directly modify
expectations for many of the modes to be studied, but the way that such a
particle interacts with the quarks and charged leptons can be used to
infer something about coupling constants associated with such interactions.
The phenomenology that is possible using data from \superb is far richer
in terms of understanding flavour couplings, than is possible at the energy frontier
machines, whereas the latter excel when it comes to direct probes of NP.  
Given the centre of mass at \superb is of the order of either 
3.8 or 10.6\gev, 
it is not possible to directly produce high energy particles in this experiment,
however it should be noted that the indirect sensitivity of many processes goes
up to $\sim 100$ TeV.  In contrast the LHC is capable of directly probing up 
to energies of $\sim 1$ TeV.  To complement its indirect search capability,
\superb will be able to make direct searches for light Higgs and Dark Matter 
particles (with masses below $10\gevcc$) that would be unobservable at the LHC.  In all of these respects
the physics programme of the LHC and \superb complement each other greatly
in the search for a deep understanding of new physics.

The path to enlightenment taken will depend on the outcome of a set of measurements
rather than by a single channel.  As a result we are faced with a response matrix of 
measurements versus new physics scenarios.  
The reason for this is that a priori we do not know which model of NP best describes
nature, and so one must look at both positive and negative signatures of a given 
model in order to identify or reject it.  The collection of observables and models
forms a {\em golden matrix}, a subset of which is shown in Table~\ref{tbl:goldenmatrix}. While any existing new 
physics scenarios can be considered as part of this matrix, these are
limited to a few specific benchmark examples to illustrate the process.

\def\onestar{\ensuremath{\star}}
\def\twostar{\ensuremath{\green\star\star}}
\def\threestar{\ensuremath{\red\star\star\star}}

\begin{table}
\caption{Golden matrix of some of the observables/modes that can be measured at SuperB.
The effect of a given model is indicated by the number of stars: \threestar, \twostar, \onestar.
The more stars the larger the effect.  Entries with $\dag$ indicate that precision 
measurement of CKM is required.  This table has been compiled based on Refs.~\cite{physicswp,Meadows:2011bk} 
and \cite{Altmannshofer:2009ne}.}
\label{tbl:goldenmatrix}
{\scriptsize
\begin{center}
\resizebox{0.75\textheight}{!}{
\begin{tabular}{|l|c|c|c|c|c|c|c|}\hline
Observable/mode & $H^+$            & MFV & non-MFV & NP           & Right-handed  & LTH & SUSY\\
                & high $\tan\beta$ &     &         & $Z$ penguins & currents      &     & \\
\hline
$\tau\to \mu\gamma$                  & & & & & & & \threestar  \\
$\tau\to \ell\ell\ell$               & & & & & & \threestar &  \\
\hline
$B\to \tau \nu, \mu\nu$              & \threestar$\dag$ & & & & & & \\
$B\to K^{(*)+}\nu \overline{\nu}$    & & & \onestar & \threestar & & & \onestar \\
$S$ in $B\to \KS\pi^0\gamma$         & & & & & \threestar & & \\
$\beta$                              & & & \threestar$\dag$ & & \threestar & & \threestar \\
$A_{CP}(B\to X_s \gamma)$            & & & \threestar & & \twostar & & \threestar \\
$BR(B\to X_s \gamma)$                & & \threestar & \onestar & & \onestar & & \\
$BR(B\to X_s \ell \ell)$             & & & \onestar & \onestar & \onestar & &  \\
$B\to K^{(*)} \ell \ell$   & & & & & & & \threestar \\
$\,\,\,$(FB Asym) & & & & & & &\\
\hline
$a_{sl}$ ($B_s\to D^{(*)}\ell\nu$)          & & & & & & \threestar  &\\
\hline
Charm mixing                         & & & & & & & \threestar  \\
CPV in Charm                         & \twostar & & & &&& \threestar \\
\hline
\end{tabular}
}
\end{center}
}
\end{table}

\subsection{The new physics energy scale: $\Lambda_{NP}$ and mass insertions}
\label{interplay:NPlagrangian:massinsertions}
$\,$
As briefly mentioned in Section~\ref{bphysics}, flavour observables can be
used to infer the energy scale of NP in different models. In terms of 
a general SUSY scenario with mass insertions (MI), one has a set of squark mixing matrices
for interactions with different helicities (left or right-handed).  These
are analogues of the CKM matrix where the
off-diagonal terms describe transitions from the $i^{th}$
to the $j^{th}$ generation generation of squark.  These are parameterised
by $(\Delta_{ij})_{kl}$, where the $k,l = L, R$ indices denote which combination of
left or right handed interactions are described.  In general one
can constrain a number of the $(\delta_{ij})_{LR}$ parameters using flavour 
observables where
\begin{eqnarray}
(\delta_{ij})_{LR} = (\Delta_{ij})_{LR} / \Lambda_{NP}.
\end{eqnarray}
Here the parameters $\delta$ are simply constrained to be less than one.
This raises an interesting point related to the 
use of results from the intensity and high energy frontier experiments, as illustrated
by the following example.

\begin{figure}[!ht]
\begin{center}
  \includegraphics[width=10cm]%
        {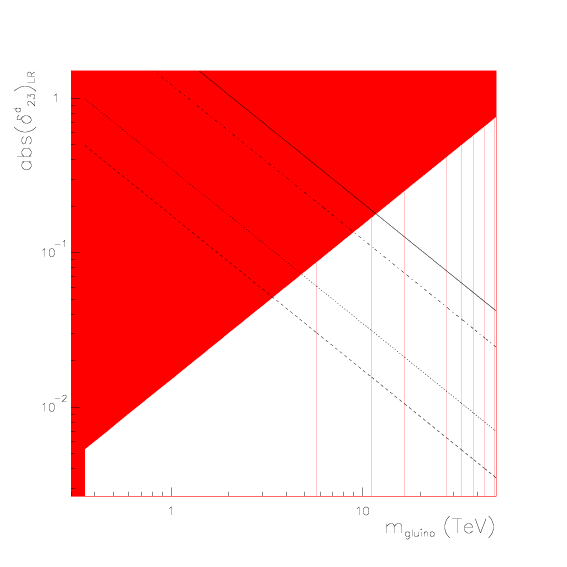}
  \caption{The constraint on the $(\delta_{23})_{LR}$ $-$ gluino mass plane obtained
  using the MI hypothesis in SUSY (figure from Ref~\cite{Bona:2007qt}).}  \label{fig:mi}
\end{center}
\end{figure}

One can combine inclusive measurements of the branching fractions of $b\to s\gamma$
and $b\to s\ell^+\ell^-$ with the direct \CP asymmetry in $b\to s\gamma$
decays at \superb to constrain the complex MI parameter $(\delta_{23})_{LR}$
as discussed in Refs~\cite{Hall1986415,Ciuchini:2002uv}.  If one assumes squark and 
gluino masses are similar, then one can relate the magnitude of this coupling to
the SUSY mass scale in a straight forward way as shown in Figure~\ref{fig:mi}.
Light SUSY has been ruled out by the LHC, which as one can see from the figure,
implies a non-trivial value for $(\delta_{23})_{LR}$.  The LHC should be able 
to probe up to masses of a few TeV by the end of this decade.  If $\Lambda_{NP}$
is ultimately fixed to a given value by a direct discovery on this time-scale, the combination of 
flavour observables from \superb can be used to make a precision measurement
of the real and imaginary parts of $(\delta_{23})_{LR}$ and teach us some of 
the details of the corresponding model.  If however the \lhc
fails to find SUSY, the same combination of flavour observables places an orthogonal
constraint on the $(\delta_{23})_{LR}-m_{\tilde{g}}$ plane, and in effect would place
an upper limit on $\Lambda_{NP}$.   For example a 5\% $(\delta_{23})_{LR}$ constraint
bounds $\Lambda_{NP}\leq 3.5$TeV.  In such a scenario, results from \superb
could be used as a guide the physics programme of the general purpose \lhc upgrade experiments,
in terms of data samples required to have sufficient energy reach for a direct discovery.
This interplay requires both high energy, and 
high intensity inputs in order to obtain the maximal level of information to
understand the model.  One can typically access scales of $\Lambda_{NP}\sim 10$TeV
using flavour observables in this scenario.

\subsection{Precision SM constraints}
\label{interplay:precisionCKM}

\begin{figure}[!ht]
\begin{center}
\resizebox{12.cm}{!}{
\includegraphics{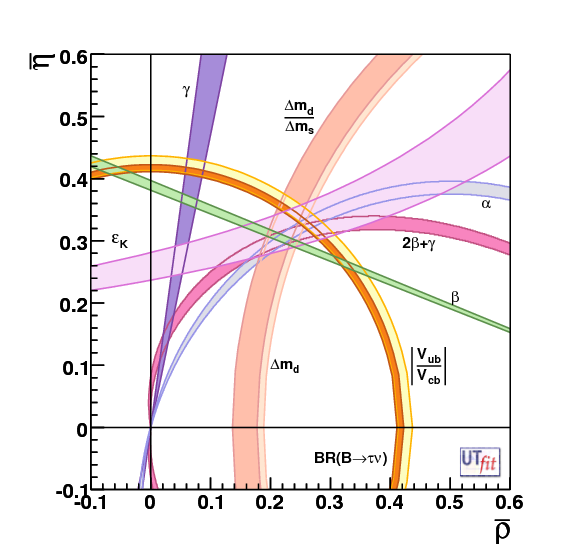}
}
\caption{An extrapolation of existing direct and indirect constraints on the CKM
mechanism to the precisions expected from \superb from Ref.~\cite{physicswp},
resulting in the so called `dream scenario' where the constraints do not coincide 
at a common point.}\label{fig:interplay:ckm}
\end{center}
\end{figure}

In 1972 Kobayashi and Maskawa~\cite{Kobayashi:1973fv} extended Cabibbo's quark-mixing model 
to three generations.  On doing this they realised that \CP violation could be naturally
introduced into theoretical descriptions of particle physics.  The experimental 
confirmation of this extended theory, i.e. the CKM mechanism, by the \babar\ experiment 
at SLAC in the USA, and the \belle\ experiment at the KEK laboratory in Japan, resulted
in Kobayashi and Maskawa being awarded a Nobel Prize in 2008.  This experimental
determination was the the completion of a set of direct tests of the CKM mechanism
with a precision of about 10\%.  The tests were measurements of the angles
$\alpha$, $\beta$, and $\gamma$ of the unitarity triangle, and they were complemented 
by a number of indirect tests.   This work established that the leading order 
contribution to \CP violation in the quark sector of the SM is a result of the 
CKM matrix but can not rule out NP effects below this level.  
The observed amount of \CP violation in the SM is not sufficient  
to describe the required matter-antimatter asymmetry in the Universe, motivating
new sources of \CP violation in quark or lepton sectors.
Indeed there are a number of models that 
can accommodate generic NP contributions that would affect the SM picture of the 
Unitarity Triangle.  \sffs
will be able to over constrain the CKM matrix through both direct and indirect measurements,
to a precision of about 1\% as indicated in Fig.~\ref{fig:interplay:ckm} (from Ref.~\cite{physicswp}).
Trivial extensions of the SM that one can test in a straightforward way, includes SM4: models where a
fourth generation of fermions are introduced to the SM.  On doing this one introduces five new
parameters to the CKM matrix, two of which are additional \CP violating phases.
The \superb and \belletwo experiments are the most 
versatile of all of the existing and proposed flavour physics experiments for 
performing such a test.  
Looking at current data, there are several tensions between
measured observables at the level of $2.5\sigma$, and it is impossible to 
simultaneously bring all of these constraints into agreement with each other
and the SM expectation~\cite{Bona:2005eu,bonaEPS2011}.  These observables are $\sin 2\beta$,
\Vub, and the measured branching fraction of the rare decay $B\to\tau\nu$.
Furthermore, there are discrepancies between inclusive and exclusive measurements
of \Vub and \Vcb that can only be investigated experimentally to a higher precision at a \sff.
While \lhcb is expected to improve the precision of our knowledge of $\sin 2\beta$
from the current precision of $0.8^\circ$ to $0.5^\circ$, 
unfortunately that experiment will not be able to study these other problematic
observables as they are all final states containing a neutrino. 
Thus a \sff is needed to resolve if these discrepancies 
are a first manifestation of NP or simply the result of statistical fluctuations.

\superb will be able to perform a precision test of the electroweak sector 
through the measurement of $\sin^2 \theta_W$ at a centre of mass energy of
10.58\gev.  The precision with which this measurement can be made is comparable
to the LEP/SLC constraint.
At \superb the $\epem\to b\overline{b}$ measurement is in a region free from hadronic uncertainties 
related to $b$ fragmentation unlike measurements at the $Z$ pole.  This measurement will
feed into the the precision electroweak fits that are currently used to predict 
the Higgs mass in the framework of the SM. Thus the \superb measurement will 
feed into both Higgs searches, and any subsequent attempts to interpret
Higgs candidates found at the LHC.

%% file: summary.tex
\section{Summary}
\label{summary}

The great challenges in fundamental physics of today range from understanding 
the evolution of the Universe from the Big Bang to the present day through 
the study of sub-atomic particles and forces at play during that time.  
New effects are expected to be uncovered while studying energy densities that 
existed in the fleeting moments after the Big Bang, and these may be 
related to our understanding of matter-antimatter asymmetries, Dark
Matter, and the existence of unknown particles. In order to improve 
our understanding of nature these different issues should not be
treated as disparate strands each with their own distinct motivation, but 
rather as distinct constraints on nature, that when combined may yield
a more lucid view of the laws of both particle physics and of the Universe.
By understanding any of these issues at a deeper level in nature, physics 
would take one step toward a Grand Unified Theory.

This process starts with the need to reconstruct a viable Lagrangian for
new physics that not only includes the physical particles that are the main
building blocks of the Universe, but also the underlying rules defining 
the behavior of the underlying forces, and hence how 
these building blocks interact.  Many ingredients will be required to start 
reconstructing the new physics Lagrangian and these will come from a number of different 
experiments, including terrestrial and satellite based astronomy, 
intensity frontier experiments like \superb, and energy frontier experiments at the Tevatron and LHC.
There are a number of precision flavour physics experiments taking data,
or in construction around the world that will also play a role in elucidating 
the structure of physics beyond the SM.

\superb is one of these experiments and it will provide many of the necessary 
ingredients required to start reconstructing the detailed texture of new physics,
and in turn will play a vital role in advancing us toward a higher 
theory of nature.  Physics topics that will be studied at \superb
range from flavour changing quark interactions, and searches for new 
phenomena like charged lepton flavour violation and \CP violation in the 
lepton sector, to searches for Dark Matter candidates and indirect searches for new particles at energy 
scales far beyond the reach of the Tevatron and the LHC.  These
topics cover many aspects of physics, and have implications in areas
beyond particle physics.  
It is interesting to note that this conclusion 
holds true irrespective of the findings of \superb: if no deviations from
the Standard Model are found, the structure of the Lagrangian is strongly
constrained, which is also the case if clear discoveries of new physics 
were to be made. In this sense \superb is a discovery experiment $-$ not necessarily 
discovery of new particles, but discovery of fragments of the underlying structure of the theory.  
Consequently, in addition to placing stringent constraints
on the type of new physics that could be possible in nature, measurements
from \superb will be able to perform many precision tests of the 
Standard Model of particle physics.

In summary the \superb experiment could start taking data as early as 2016, and 
make a diverse set of measurements of flavour related observables.
It is expected that \superb will have recorded 75\invfb in five years of 
nominal data taking, 
corresponding to 75 billion $B$, 90 billion $D$, and 70 billion $\tau$ pairs for analysis.  The 
sensitivities that one will be able to reach with such data samples can be as 
low as ${\cal O}(few \times 10^{-10})$ for clean processes.
It is worth noting that there are also important measurements that will be made
at \superb that impact the physics reach of other experiments.  Two
examples of such a measurement is the strong phase as a function of position
in the Dalitz plot for decays such as $\Dz\to \KS\pi^+\pi^-$ that feed 
into \D mixing and $\gamma$ measurements, and the branching fraction of 
$D^0\to \gamma\gamma$, which is required in order to disentangle long distance
SM contributions from NP enhancements of $D^0\to \mu^+\mu^-$.  The precision
CKM measurements that will be performed at \superb can be used to open over-constrain
the SM, and if one finds measurement consistent with theory, then this also 
provides the gateway for NP searches at other experiments. For example the rare kaon decay
experiments searching for $K\to \pi\nu\overline{\nu}$ will be able to use their results
to test the SM, however if CKM uncertainties are reduced, one could improve the 
sensitivity of those channels to NP.  Ref.~\cite{Meadows:2011bk} contains a succinct summary 
of the expected precisions obtainable for the core measurements to be made at \superb.

%% file: ack.tex
The author would like to acknowledge the exhaustive efforts of \superb experimental 
and theoretical communities over the past decade.  These efforts would not have
existed without the creativity of an accelerator team who have repeatedly inspired 
the experimentalists to rise to the challenge of understanding the potential 
of ever larger data samples.  This report is largely based on the work presented
in the \superb progress reports of Ref.~\cite{physicswp,acceleratorwp,detectorwp}
and the more recent update of Ref.~\cite{Meadows:2011bk}.